\documentclass[12pt,a4paper,final]{iopart}

\usepackage{iopams}  
\expandafter\let\csname equation*\endcsname\relax
\expandafter\let\csname endequation*\endcsname\relax
\usepackage{amsmath,amssymb}

\usepackage{graphicx}
\usepackage[breaklinks=true,colorlinks=true,linkcolor=blue,urlcolor=blue,citecolor=blue,pagebackref=true]{hyperref}
\usepackage{color}

\usepackage[caption=false]{subfig}

\newcommand{\integral}[3]{\int_{#2}^{#3} \!\! \mathrm{d} #1 \,}
\newcommand{\difffrac}[2]{\frac{\mathrm{d} #1}{\mathrm{d} #2}}

\newcommand{\ee}[1]{\rme^{#1}} 
\newcommand{\ii}{\rmi}

\newcommand{\ket}[1]{\left| {#1} \right\rangle}

\newcommand{\ketbra}[2]{\left| {#1} \right\rangle\!\left\langle {#2} \right|}

\newcommand{\intHd}[1]{H_{I,\,#1}}
\newcommand{\intH}{H_{I}}

\newcommand{\id}{\mathbb{I}}

\newcommand{\exptval}[1]{\left\langle {#1} \right\rangle}
\newcommand{\comm}[2]{\left[{#1},{#2}\right]}

\renewcommand{\Re}{\mathrm{Re}} 
\renewcommand{\Im}{\mathrm{Im}} 

\begin{document}

\title{Quantum Signaling in Relativistic Motion and Across Acceleration Horizons}

\author{Robert H. Jonsson$^{1,2}$}
\address{$^1$Microtechnology and Nanoscience, MC2, Chalmers University of Technology, SE-412 96 G\"oteborg, Sweden}
\address{$^2$Department of Applied Mathematics, University of Waterloo, Waterloo, Ontario, N2L 3G1, Canada}
\ead{\mailto{robejon@chalmers.se} \mailto{rhjonsso@uwaterloo.ca}}

\begin{abstract}
The quantum channel between two particle detectors provides a prototype framework for the study of wireless quantum communication via relativistic quantum fields.
In this article we calculate the classical channel capacity between two Unruh-DeWitt detectors arising from couplings within the perturbative regime. To this end, we identify the detector states which achieve maximal signal strength.
We use these results to investigate the impact of relativistic effects on signaling between detectors in inertial and uniformly accelerated motion which communicate via a massless field in Minkowski spacetime.
\end{abstract}


\section{Introduction}

Our ability to process information and transmit it across spacetime is impacted both by spacetime curvature and by quantum effects. 
This interplay of general relativity, quantum theory and  information theory has important implications from various points of view.
On the one hand, from a technological perspective,   it determines ultimate limits on information technology that arise from the fundamental laws of nature. Furthermore, the  combination of hitherto unexplored phenomena can lead the way to novel methods of information processing.
On the other hand,  the information theoretic approach has added a very fruitful perspective, and a deeper understanding  of the fundamental interplay between gravity and quantum theory.
This is illustrated by the continuing vibrant research activity around, e.g., the black hole information paradox or the holographic principle. 

Recent examples,  that are of particular relevance to this article, include an ultimate limit on the capacity of quantum communication channels that arises from entropy bounds on quantum fields \cite{bousso_universal_2016}. Also, it has been shown that the spacetime curvature induced by the energy necessary to operate a quantum measurement device, limits the very precision  with which spacetime geometry can be measured \cite{lloyd_quantum_2012}. 
Relativistic (quantum) bit commitment and summoning \cite{kent_unconditionally_1999,kent_quantum_2012,hayden_summoning_2012} are examples of an information processing task whose implementation relies  both on the laws of relativity and quantum theory.

This article addresses the question how   communication via relativistic quantum fields is impacted by relativistic motion of the signaling devices. 
To obtain a prototype framework which allows to study the combination of relativistic and quantum effects, we model the signaling devices as basic, first-quantized quantum systems. We assume that the devices move through spacetime along the classical wordlines which describe the sender's and receiver's motion. Along these wordlines the devices ccouple to the quantum field to emit and receive signals.

To model the interaction between signaling device and quantum field we employ the renown Unruh-DeWitt particle detector model, which provides a simple model for the interaction between an atom and a background quantum field.
The Unruh-DeWitt  particle detector was introduced in the study of quantum field theory in curved spacetimes \cite{unruh_notes_1976,hawking_quantum_1979}. In curved spacetimes two observers, in general, do not agree on the particle content, or even the presence of particles, in a given state of a quantum field.  
In the particle detector model the excitation of the model atom through the interaction with the quantum field, is interpreted as the detection of a particle with respect to an observer moving along the atom's worldline.
In this way, the particle detector model offers an operational approach to  the phenomena that spacetime curvature causes in quantum fields, such as the Unruh effect, Hawking radiation, or particle creation in expanding universes \cite{unruh_notes_1976,gibbons_cosmological_1977}.



Consequently, particle detector models have become an important tool in relativistic quantum information to explore the quantum information theoretical properties and potential of relativistic quantum fields. 
Here an influential result was, e.g., the rediscovery \cite{reznik_violating_2005} of the fact that entanglement present in the vacuum state of a field can be used to entangle  detectors, even if they are spacelike separated \cite{summers_bells_1987,summers_vacuum_1985}. Subsequent works analyzed how this effect is impacted by spacetime geometry and relativistic motion \cite{martin-martinez_entanglement_2014,steeg_entangling_2009,salton_acceleration-assisted_2015,cliche_vacuum_2011}. 
Another fascinating phenomenon that uses the entanglement present in the field, combined with classical communication, is quantum energy teleportation. 
Here one party performs a measurement on the field and sends the outcome to another party, which  then uses this information  to extract energy from the quantum fluctuations of the field \cite{hotta_quantum_2008,hotta_controlled_2010,hotta_quantum_2014,verdon-akzam_asymptotically_2016}.

In this article we aim to study wireless quantum communication on a fundamental level. Therefore, we use the Unruh-DeWitt particle detector as a prototype model for the interaction of relativistic observers with quantum fields. 
For this, we equip the sender and receiver  with particle detectors,  as basic quantum devices, which they use to transmit signals between each other through a relativistic quantum field.
The influence of the sender's initial detector state on the receiver's final state constitutes, in quantum information theoretical terms,  a quantum channel \cite{nielsen_quantum_2010}.
Therefore, studying communication in this  framework, all methods and results on quantum channels from quantum information theory can be employed to study the combined impact of relativistic and quantum effects. In fact, they can be quantified in terms of their effects on the channel capacity.
This approach was first proposed   in \cite{cliche_relativistic_2010}. It has since been  used to highlight interesting, and potentially rather counter-intuitive, features of information propagation in massless  fields.
Whereas signals in massless fields often are thought of as being carried by field quanta that propagate strictly at the speed of light, and which carry energy from the sender to the receiver, it was shown that signals in massless fields can propagate slower than light, and can transmit information without carrying any energy from the sender to the receiver \cite{jonsson_information_2015,jonsson_information_2016}.

Only in spacetimes where Huygens' principle applies do timelike signals not occur. This is the case in 3+1-dimensional Minkowski spacetime. However, in lower dimensions, and in general curved spacetimes, Huygens' principle generally does not apply \cite{czapor_hadamards_2007}.
For example, timelike signals in massless fields occur  in expanding universes and could be of interest for obtaining information about early universe cosmology \cite{blasco_violation_2015,blasco_timelike_2016,simidzija_information_2017}.

\subsection{Results and Structure of Article}

In this article we study the impact of inertial and accelerated motion on the signal strength between detectors in Minkowski spacetime, and we identify a characteristic difference between null and timelike signals in massless fields.

For null signals, we find that the signal strength is maximized if the sender's detector and the receiver's detector are resonant, as one would intuitively expect. If the two parties are in relative intertial motion to each other, the detector energy gaps need to be detuned so as to account for the relativistic Doppler shift (Section \ref{sec:inertial}.). 

However, if one of the parties is uniformly accelerated, the Doppler shift grows infinitely at early and late times. This results in a   bound on the signal strength between an inertial and an accelerated observer that are separated by the acceleration horizon. (Section \ref{sec:accelerated}.) Even if they have an infinite amount of time to interact with the field to send and receive the signal, the signal strength does not grow above a certain limit.

Timelike signals are set apart from null signals by a particular property: They do not require resonance or synchronization between the sender and receiver to maximize the signal strength. (Section \ref{sec:timelike}.)
The receiver's optimal choice of coupling parameters and coupling times  is independent of the sender's coupling parameters or worldline. The receiver can just individually and locally optimize their own coupling parameters in order to optimally detect any  signal that a sender may have emitted into their future lightcone.
However, similar to the signal strength across the acceleration horizon, the signal strength of timelike signals is bounded. It  cannot be increased by allowing the sender and receiver interact with the field for longer times. 

We also demonstrate that the decoupling of information transmission from energy transmission, that was demonstrated for timelike signals \cite{jonsson_information_2015,jonsson_information_2016}, is  a general property of signals encoded into the amplitude of massless fields.
When the distance between sender and receiver is increased  the signal strength  decays with a lower power of the distance than the energy content the signal carries from the sender to the receiever. This holds both for timelike and null signals.  (Section \ref{sec:rest}.)

We use time-dependent perturbation theory to describe the interaction of the detectors with the field, which is the most common approach in the literature. While this puts no restrictions on the detectors' worldlines and parameters, it means that overall the interaction is considered to be perturbatively small. 

We analyze which classical channel capacity, as captured by different capacity measures, arises from the leading order signaling effects (in Section \ref{sec:capacity}). And we identify the initial states in which sender and receiver need to prepare their detectors to achieve the optimal signal strength. (Section \ref{sec:bloch}.)

The key result here is that the leading order signal strength is captured by a comparatively simple expression.   
It should be feasible to evaluate this expression in many relativistic communication scenarios of interest, beyond the scenarios in Minkowski spacetime treated in this article.

This result was anticipated in \cite{jonsson_decoupling_2016}, but now in Section \ref{sec:generalinput}, we show that the expression for the leading order signal strength is indeed general and optimal for all possible initial states of the detectors.
In view of this result, Sections \ref{sec:capacity} and \ref{sec:minkowski} give a self-contained discussion of results which were first derived in  \cite{jonsson_decoupling_2016}. The interested reader may find more details, e.g., on calculations there.


Throughout the article we use natural units $\hbar=c=1$. 

\section{A prototype model of wireless quantum communication}\label{sec:channel}

In the following we introduce the Unruh-DeWitt interaction Hamiltonian, and discuss the structure of the qubit quantum channel between two detectors. Before entering into technical detail, we  review the general idea of the framework.

Our aim is to study the fundamentals of wireless quantum communication via quantum fields between relativistic observers. To this end, the sender, called Alice, and the receiver, called Bob, are equipped with simple (first-quantized) quantum systems as communication devices. Here we will just use  two-level sytems, i.e., qubits. 
The communication devices can interact with the quantum field locally. We assume that Alice and Bob can control the interaction by switching the coupling on and off.
The quantum field serves as the medium which carries the signals from Alice to Bob.

If Alice is to send a message to Bob she encodes the message into the initial state of her qubit, before her qubit and the quantum field interact. Then, to emit the  signal, she couples her qubit to the field for some time. The interaction between Alice's qubit and the field cause disturbances to the field amplitude that propagate through spacetime and reach Bob.

To record Alice's message, Bob couples his device to the field such that Alice's signal modulates the interaction between the field and Bob's device.
We assume that Alice and Bob can control the coupling between their device and the field. However, we assume that they can perform measurements only on their signaling devices. Therefore, Bob  has to infer Alice's message from a measurement only on the final state of his device, after decoupling his device from the field.

How much information Alice can transmit to Bob  depends on how much influence Alice's choice of initial state has on Bob's final state. 
Both relativistic and quantum effects will impact this influence of Alice on Bob. And the size of their combined impact can be quantified in terms of the information-theoretical channel capacity that arises from Alice's influence on Bob. For this a  wide range of results and methods from quantum information theory are available, because the  map from Alice's initial state to Bob's final state is a quantum channel map \cite{nielsen_quantum_2010}.

\subsection{Unruh-DeWitt particle detector model}

To model the communication devices and their interaction with the quantum field we use the Unruh-DeWitt particle model \cite{unruh_notes_1976,hawking_quantum_1979}.
An Unruh-DeWitt particle detector consists of a two-level system whose energy eigenstates we denote by $\ket{e_d},\ket{g_d}$. The energy gap of the detector is $\Omega_d$, i.e., the Hamiltonian of the free, uncoupled detector reads $H_d=\Omega_d\ketbra{e}{e}$. The detector moves through spacetime along a classical wordline $x_d( \tau_d)$, which can be parametrized by the detector's proper time $\tau_d$.

Along its wordline the detector interacts with a scalar real Klein-Gordon field $\phi(x)$. In the Dirac interaction picture the interaction Hamiltonian reads
\begin{eqnarray}\label{eq:udwinth}
\intHd{d}(\tau_d) &= \lambda_d \,\eta_d(\tau_d) \left( \ee{\ii \Omega_d \tau_d}\ketbra{e_d}{g_d}+ \ee{-\ii \Omega_d \tau_d} \ketbra{g_d}{e_d}\right) \otimes \phi(x_d(\tau_d)).
\end{eqnarray}
Here $\eta_d(\tau_d)\in[0,1]$ is a real-valued switching function which determines at what time the detector and the field are interacting.

The coupling constant $\lambda_d$   sets the strength of the interaction between the detector and the field. We  use the coupling constant as the perturbative parameter in the perturbative treatment of the time evolution, as is commonplace in the literature.
However, it is worth noting  that $\lambda_d$ generally is  dimensionful. It is dimensionless only in 3+1-dimensional spacetimes. In $n+1$-dimensional spacetime it has mass dimension $\left[\lambda_d\right]=(3-n)/2$.  For example, in 1+1 dimensions $[\lambda_d]=1$, i.e., it corresponds to a mass or energy. In order for the perturbative analysis to be valid, $\lambda_d$ then needs to be small with respect to some energy scale, which can be set by the detector energy gap $\Omega_d$ (see Section \ref{sec:inertial}) or the detector's acceleration (see Section \ref{sec:accelerated}).

\begin{figure}
\centering
\includegraphics[width=0.45\textwidth]{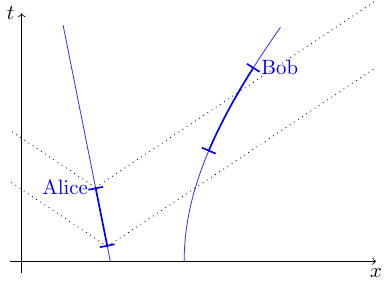}
  \caption{ Spacetime diagram of a signaling scenario. Alice and Bob move through spacetime along independent worldlines. They couple their communication devices (Unruh-DeWitt particle detectors) to the field during finite time windows which are determined by the detector switching functions $\eta_d(\tau_d)$ in \eqref{eq:udwinth}. The dotted lines indicate lightrays emanating from Alice.}
  \label{fig:spacetimediag}
\end{figure}

As sketched in Figure \ref{fig:spacetimediag}, we are interested in scenarios where Alice and Bob move independently through spacetime. This means that their individual proper times can evolve very differently. In such scenarios it can be more convenient to express the interaction Hamiltonian as  $\intH(t)=\difffrac{\tau}{t} \intH(\tau) $, where $t$ is a global time coordinate. It then generates time translations with respect to the coordinate time, rather than with respect to the detector proper time. (For a detailed discussion of this aspect see, e.g.,  \cite{brown_detectors_2013}.)
The total interaction Hamiltonian for a scenario as in Figure \ref{fig:spacetimediag} then reads
\begin{eqnarray}
\fl\quad\intH(t)=\sum_{d=A,B} \lambda_d \,\eta_d(\tau_d(t)) \difffrac{\tau_d(t)}{t} \left( \ee{\ii \Omega_d \tau_d}(t)\ketbra{e_d}{g_d}+ \ee{-\ii \Omega_d \tau_d(t)} \ketbra{g_d}{e_d}\right) \otimes \phi(x_d(t)).
\end{eqnarray}
Now, the detectors' worldlines are parametrized by $t$, the coordinate time. The choice of switching function determines when Alice and Bob interact with the field  along their worldlines.

\subsection{The Quantum Channel between two Detectors}
We assume that the detectors and the field start out in a product state, before Alice and Bob interact with the field. We assume that the field begins in the vacuum state (generalisations to other states diagonal in the Fock basis are straightforward) and denote the initial state of the total system by
\begin{eqnarray}\label{eq:initialstate}
\rho_0=\rho_{0,A}\otimes \rho_{B,0} \otimes \ketbra00 = \begin{pmatrix} \theta & \gamma \\ \gamma^*&\beta\end{pmatrix}\otimes \begin{pmatrix} \varphi&\delta \\ \delta^* & \kappa\end{pmatrix}\otimes \ketbra00.
\end{eqnarray}
Under the interaction between the detectors and the field, the system evolves unitarily into a final state $\rho= U \rho_0 U^\dagger$ in which generally the field and detectors are entangled. To obtain  $\rho_B$, the  final state  of Bob's detector, i.e., the state from which Bob tries to retrieve Alice's message, we take the partial trace over the field and Alice's detector.
\begin{eqnarray}
\rho_B&=\tr_{A,F} \rho =\tr_{A,F} U \rho_0 U^\dagger
\end{eqnarray}

We are interested in how much information Alice is able to transmit to Bob by means of their interaction with the quantum field. This is determined by how Bob's final state depends on Alice's choice of input state. This dependency is given by the quantum channel map $\xi$ from Alice's initial state to Bob's final state
\begin{eqnarray}
\xi: \rho_{A,0}\mapsto \rho_B = \tr_{A,F} U \rho_0 U^\dagger.
\end{eqnarray}
Therefore, the amount of information that Alice is able to transmit to Bob can be quantified in terms of the information theoretical capacity of $\xi$. 
As we will show in this paper,  within the perturbative regime the leading order contributions to the signal strength and the classical channel capacity take a particularly simple form.

The general structure of Bob's final state, resulting from  Alice's and Bob's interaction with the field, as derived and detailed in  \cite{jonsson_quantum_2014,cliche_relativistic_2010}, is
\begin{multline}\label{eq:chanstructgeneral}
\rho_B=\begin{pmatrix}\varphi & \delta \\ \delta^* & \kappa\end{pmatrix}+\begin{pmatrix}\kappa P + \varphi Q & \delta R + \delta^* S^* \\ \delta^* R^* +\delta S & -\kappa P -\varphi Q\end{pmatrix}\\
+ \gamma \begin{pmatrix} \delta I+\delta^* J & \kappa C+ \varphi G \\ \kappa D + \varphi H & -\delta I-\delta^* J  \end{pmatrix} + \gamma^* \begin{pmatrix}\delta J^* + \delta^* I^* & \kappa D^* + \varphi H^*\\ \kappa C^*+\varphi G^* &  -\delta J^* -\delta^* I^* \end{pmatrix}\\
+\theta \begin{pmatrix} \kappa A+\varphi E & \delta K + \delta^* L^*\\ \delta L+\delta^* K^* & -\kappa A - \varphi E
\end{pmatrix}+\beta \begin{pmatrix} \kappa B+\varphi F & \delta M + \delta^* N^* \\ \delta N + \delta^* M^* & -\kappa B - \varphi F \end{pmatrix},
\end{multline}
where capital Latin letters indicate coefficients that are determined by all the physical parameters of the scenario, i.e., detector energy gaps, switching functions and detector wordlines.

The contribution to Bob's final state in the first line is independent of Alice's presence. In fact, it is the sum of Bob's initial state and a term arising solely from the interaction of Bob's detector with the field.
The contributions in the  second and third line modulate Bob's final state depending on Alice's choice of initial state and, thus, are relevant for information transmission from Alice to Bob.

When the interaction between the detectors and the field is treated perturbatively, the time evolution operator is expanded using the Dyson series
\begin{eqnarray}\label{eq:dysonseries}
\fl\quad U=\mathcal{T}\exp\left(-\ii\intH\right)=\id-\ii\integral{t_1}{}{}\intH(t_1)-\integral{t_2}{}{}\integral{t_1}{}{t_2}\intH(t_2)\intH(t_1)+... \quad.
\end{eqnarray}
This results in a perturbative expansion of the channel coefficients and accordingly of Bob's final state \eqref{eq:chanstructgeneral} \cite{jonsson_quantum_2014}. The leading order contributions in the expansion of Bob's final state are
\begin{multline}\label{eq:chanleadingorder}
\rho_B = \begin{pmatrix} \varphi& \delta\\ \delta^* & \kappa\end{pmatrix} + \begin{pmatrix} \kappa P_2+\varphi Q_2 & \delta R_2+ \delta^* S_2^*\\ \delta^* R^*_2+\delta S_2 & -\kappa P_2 -\varphi Q_2\end{pmatrix} \\
+\left[ \gamma \begin{pmatrix} \delta D_2+\delta^* C_2 & (\kappa-\varphi) C_2\\ (\kappa-\varphi) D_2 & - \delta D_2-\delta^* C_2\end{pmatrix}+ H.c.\right] +\mathcal{O}(\lambda^4),
\end{multline}
where, in particular,
\begin{eqnarray}
\fl \quad&C_2= \lambda_A \lambda_B\integral{t_1}{}{}\integral{t_2}{}{t_1} \chi_A(t_2) \chi_B(t_1) \ee{\ii (\Omega_B\tau_B(t_1)- \Omega_A\tau_A(t_2) )} \comm{\phi(x_A(t_2))}{\phi(x_B(t_1))} \label{eq:c2}\\
\fl\quad&D_2=- \lambda_A \lambda_B\integral{t_1}{}{}\integral{t_2}{}{t_1} \chi_A(t_2) \chi_B(t_1) \ee{-\ii (\Omega_B\tau_B(t_1)+ \Omega_A\tau_A(t_2) )} \comm{\phi(x_A(t_2))}{\phi(x_B(t_1))} \label{eq:d2},
\end{eqnarray}
and we absorbed the switching function and the time derivative of the detector proper time into $\chi_d(t)=\eta_d(\tau_d(t)) \difffrac{\tau_d(t)}{t}$. 
We use the symbol $\mathcal{O}(\lambda^4)$ to denote terms that are proportional to the fourth or higher powers of the coupling constants, i.e., to terms that contain a factor of, e.g., $\lambda_B^4$ or $\lambda_A^2\lambda_B^2$.
(The terms $P_2,Q_2,R_2,S_2$ are given in \ref{app:terms}.) The terms $C_2, D_2$ etc., are dimensionless quantities, which is clear from their appearance in Bob's final density matrix above. They represent only perturbative corrections to Bob's state and, therefore, ought have small absolute value ($|C_2|<<1$). Larger values, e.g., due to large coupling constants or long interaction times, would cease to lie within the regime of perturbation theory.

As discussed already in \cite{jonsson_quantum_2014}, the expansion of $\rho_B$ indicates  that  Alice's optimal choice of signaling states, within the perturbative regime, need to be equal weighted superpositions of energy eigenstates. This is because  they maximize $\left| \gamma\right|$, the size of the off-diagonal entries  of Alice's initial density matrix $\rho_{A,0}$, which are the only entries of $\rho_{A,0}$ affecting $\rho_B$ at leading order $\mathcal{O}(\lambda^2)$. If Alice prepares her detector in an energy eigenstate she only affects Bob's final state at next-to-leading order $\mathcal{O}(\lambda^4)$.

In the following we show that the sum of absolute values of the leading order signaling contributions, $|C_2|+|D_2|$ can be used to measure the leading order signal strength that a specific communication scenario allows for.

\section{Bloch Sphere Representation and Optimal Initial States}\label{sec:bloch}
In this section we determine which initial states Alice and Bob need to prepare their detectors in, in order to maximize the leading order signal strength.
For this we use the Bloch sphere representation of the channel which also gives an intuitive picture of the general structure of the quantum channel at hand.

We find that the Alice's optimal choice of signaling states, which are determined by the channel coefficients $C_2$ and $D_2$,  are a pair of equal-weighted superpositions of energy eigenstates. Bob can choose his initial state from a one-parameter family of states including the detector's energy eigenstates.

The Bloch sphere representation uses that any qubit density matrix $\rho$ can be represented as a real, three-dimensional vector $\brho=\left(\rho_X, \rho_Y, \rho_Z\right)$ through \cite{nielsen_quantum_2010}
\begin{eqnarray}
\fl\quad\rho=\frac12\left(\id+\brho\cdot\bsigma\right)=\frac12\left(\id+ \rho_X \sigma_X+\rho_Y \sigma_Y+\rho_X \sigma_X\right)=\frac12\begin{pmatrix} 1+\rho_Z & \rho_X-\ii \rho_Y\\ \rho_X+\ii \rho_Y & 1-\rho_Z \end{pmatrix}
\end{eqnarray}
with the standard Pauli matrices $\sigma_X,\sigma_Y,\sigma_Z$. The Bloch vector of a pure qubit state has norm $\| \brho\| =\sqrt{\rho_X^2+\rho_Y^2+\rho_Z^2}=1$, whereas for mixed states $\| \brho\| < 1$.
In this representation, a quantum channel $\xi: \rho_i\mapsto \rho_o$ between qubits is represented by an affine map \cite{nielsen_quantum_2010},
\begin{eqnarray}
\brho_i\mapsto\brho_o= \bi{M} \brho_i +\bi{v},
\end{eqnarray}
with a real-valued 3x3-matrix $\bi M$ and a constant vector $\bi{v}$.

\subsection{Bob Initialized in Ground State}

\begin{figure}
\centering
\includegraphics[width=0.5\textwidth]{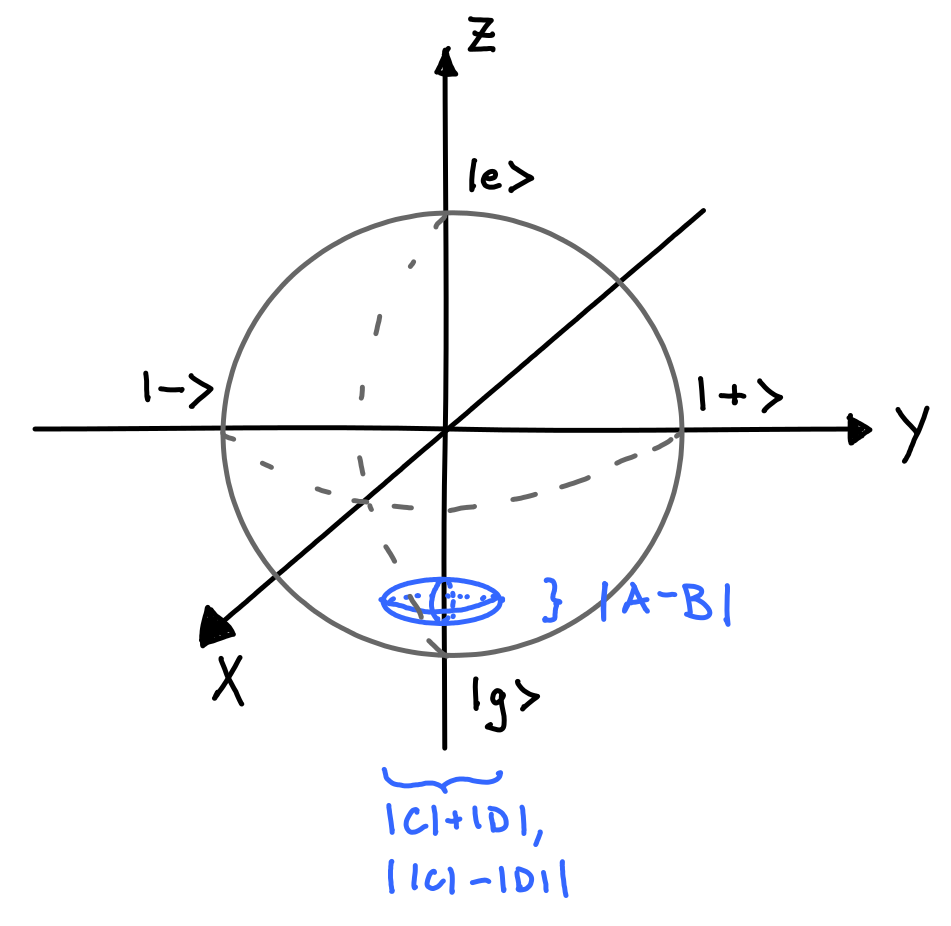}
\caption[Bloch sphere picture of channel]{Sketch of  possible final states of Bob, when he initializes his detector in the ground state $\ket{g_B}$ before the interaction. The Bloch sphere of Alice's initial states is contracted to an ellipsoid close to the receiver's ground state. 
The ellipsoid's diameter in the $X-Y$-plane is determined by the absolute values of the  channel coefficients $C$ and $D$ from equation \eqref{eq:chanstructgeneral}. 
The ellipsoid's diameter along the $Z$-axis is determined by the coefficients  $A$ and $B$. (Figure reproduced from \cite{jonsson_decoupling_2016}.)
}
\label{fig:blochspherechannel}
\end{figure}

We first apply this representation   to the channel between two particle detectors in the special case  where Bob initializes his detector in the ground state, i.e., the entries of his density matrix in \eqref{eq:initialstate} are  $\kappa=1$ and $ \varphi=\delta=0$.
Then, from \eqref{eq:chanstructgeneral}, we find that in the Bloch representation the channel maps Alice's initial state $\brho_{A,0}$ to 
\begin{eqnarray}\label{eq:BobgroundBlochchannel}
\fl\quad\brho_B&= \bi{M}\brho_{A,0}+\bi{v}\nonumber\\
\fl\quad&= \begin{pmatrix} \Re(C+D) & \Im(C+D) &0\\\Im(-C+D)&\Re(C-D)&0\\0&0&A-B\end{pmatrix} \brho_{A,0} + \begin{pmatrix} 0\\0\\2P+A+B-1\end{pmatrix}.
\end{eqnarray}

The action of the channel is most clearly seen from the singular value decomposition of the channel matrix. Before obtaining it, we rewrite the complex channel coefficients as $C=|C|\ee{\ii\phi_C}$ and $D=|D|\ee{\ii\phi_D}$. (Note that the coefficients $A$ and $B$ are real-valued \cite{jonsson_quantum_2014}.) Then we can decompose $\bi{M}$ as
\begin{eqnarray}
\bi{M} = \bi{U ODO^{t}}
\end{eqnarray}
with 
\begin{eqnarray}
\fl\quad\bi{U}=\begin{pmatrix}\cos\phi_C&\sin\phi_C&0\\-\sin\phi_C&\cos\phi_C&0\\0&0&1\end{pmatrix}, \qquad \bi{O}= \begin{pmatrix}\cos\frac{\phi_C+\phi_D}2&-\sin\frac{\phi_C+\phi_D}2&0\\ \sin\frac{\phi_C+\phi_D}2&\cos\frac{\phi_C+\phi_D}2&0\\0&0&1\end{pmatrix}
\end{eqnarray}
and the diagonal matrix
\begin{eqnarray}
\bi{D}=\begin{pmatrix} |C|+|D|&0&0\\0&|C|-|D|&0\\0&0&A-B\end{pmatrix}.
\end{eqnarray}
Therefore, as sketched in Figure \ref{fig:blochspherechannel}, the action of the channel is the following: The three-vector $\brho_{A,0}$ of Alice's initial state is first rotated in the $X-Y$-plane by the angle $(\phi_C+\phi_D)/2$ clockwise around the $Z$-axis. Then the multiplication by $\bi{M}$ reduces the length of the vector by different factors along the three axes. Consequently, the multiplication by $\bi{UO}$ corresponds to another rotation in the $X-Y$-plane by the angle $(\phi_C-\phi_D)/2$ clockwise around the $Z$-axis. Finally, the vector obtained from these operations is added to the constant vector $\bi{v}$ which lies close to the ground state in which Bob was initialized.

Which initial states should Alice use in order to be able to most efficiently transmit classical information to Bob? As we discuss in the subsequent section in more detail, Alice needs to find a pair of initial states, $\rho_{1}$ and $\rho_{2}$, which yield output states $\xi(\rho_{1})$ and $\xi(\rho_{2})$ that Bob is able to distinguish as well as possible. 
For this, the trace distance $D(\xi(\rho_1),\xi(\rho_2))$ between the output states needs to be maximized \cite{nielsen_quantum_2010}. 

In the Bloch sphere picture the trace distance between two qubit states is equal to half the Euclidean distance between the states' three-vectors:
\begin{eqnarray}
D( \brho_1,\brho_2) = \frac{\| \brho_1 - \brho_2\|}2.
\end{eqnarray}
Therefore, Alice needs to choose a pair of orthogonal pure qubit states 
because the Bloch vectors of orthogonal states point in opposite directions, i.e., $\brho_1=-\brho_2$. Furthermore, she needs to choose the states such that 
they get multiplied by the largest of the elements of the diagonal matrix $\bi{D}$.

The largest element of $\bi D$ is always $|C|+|D|$, in the perturbative regime. This is because the first two entries are of leading order in perturbation theory $|C|\pm|D|\sim |C_2|\pm |D_2|+\mathcal{O}(\lambda^4)$ whereas the third element is of next-to-leading order $A-B\sim A_4-B_4+\mathcal{O}(\lambda^6)$ only \cite{jonsson_quantum_2014}.
Alice's optimal choice of signaling states, therefore, are
\begin{eqnarray}\label{eq:aliceoptstate}
\brho_{1,2}=\pm \begin{pmatrix} \cos\frac{\phi_C+\phi_D}2\\ \sin\frac{\phi_C+\phi_D}2 \\0\end{pmatrix}
\end{eqnarray}
which is equivalent to the density matrix elements $\theta=\beta=1/2$ and $\gamma= \pm\ee{-\ii(\phi_C+\phi_D)/2}/2$ of Alice's initial state in \eqref{eq:initialstate}, i.e., the two pure states $\left(\ket{e_A}\pm\ee{\ii(\phi_C+\phi_D)/2}\ket{g_A}\right)/\sqrt2$.

For this choice of initial state the trace distance between Bob's final states is maximal. Its leading order expansion is given by the sum of the absolute value of the leading order contributions to the channel coefficients:
\begin{eqnarray}
D\left(\xi(\brho_1),\xi(\brho_2)\right) = |C|+|D| \sim |C_2|+|D_2|+\mathcal{O}(\lambda^4).
\end{eqnarray}

\subsection{General Optimal Initial States}\label{sec:generalinput}

In the following we show that this leading order perturbative behaviour of the trace distance is optimal for all possible choices of initial states of Bob. It can be achieved by a one-dimensional family of input states for Bob which form a circle on the Bloch sphere which includes the ground state $\ket{g_B}$ and the excited state $\ket{e_B}$ of Bob's detector. Alice's choice of signaling states is independent of Bob's initial state.

To show this, we look at the leading order of the Bloch sphere representation of the channel for an arbitrary initial state of Bob. From \eqref{eq:chanleadingorder} it follows that, for an arbitrary input state $\brho_{B,0}$ of Bob in \eqref{eq:initialstate}, the Bloch vector of Bob's final state is  to leading order 
\begin{eqnarray}
\fl\quad\brho_B= \bi{v}+ \underbrace{\begin{pmatrix} (\kappa-\varphi) \Re(C_2+D_2) & (\kappa-\varphi)\Im(C_2+D_2) &0\\ (\kappa-\varphi) \Im(-C_2+D_2)&(\kappa-\varphi)\Re(C_2-D_2)&0\\2\Re ( \delta(C_2^*+D_2) ) & 2\Im ( \delta^* (C_2-D_2^*)) &0 \end{pmatrix}}_{=:\bi{M}_2 \sim\mathcal{O}(\lambda^2)} \brho_{A,0} +\mathcal{O}(\lambda^4).
\end{eqnarray}
In order to achieve a maximal trace distance, Alice's initial states $\brho_{1}$ and $\brho_{2}=-\brho_{1}$ need to maximize the norm  $\| \bi{M}_2 \brho_{1}\|$ (which  is equal to $\|\bi{M}_2\brho_{2}\|$). 
For this $\brho_{1,2}$ need to lie in the $X-Y$-plane of the Bloch sphere and, accordingly, have a vanishing $Z$-component, because the third column of $\bi{M}_2$ vanishes. 
This illustrates once more that Alice's optimal choice are equal-weighted superpositions of energy eigenstates, as mentioned above.

The remaining question is  in which direction the Bloch vectors of the optimal signaling states are directed. To answer this we denote Alice's initial state by $\brho_{1,2}=\pm (\cos \alpha,\sin \alpha,0)$.   The norm $\| \bi{M}_2 \brho_1\|$ can be separated into contributions from its first two components and its $Z$-component
\begin{multline}
\left\| \bi{M}_2 \brho_1\right\|^2 = (\kappa-\varphi)^2 \left\| \begin{pmatrix} \Re(C_2+D_2) & \Im(C_2+D_2) \\ \Im(-C_2+D_2)& \Re(C_2-D_2) \end{pmatrix} \begin{pmatrix}\cos \alpha \\ \sin \alpha\end{pmatrix}\right\|^2 \\
+ 4  \left[ \cos \alpha \Re \left(\delta (C_2^*+D_2)\right) +\sin \alpha \Im \left(\delta^* (C_2-D^*_2)\right) \right]^2.
\end{multline}
The first term originating from the 2x2-matrix is identical to the upper diagonal block of the channel matrix we analyzed before in \eqref{eq:BobgroundBlochchannel}. Just as there, it is maximal when $\alpha=(\phi_{C_2}+\phi_{D_2})/2$, for $C_2=\left|C_2\right| \ee{\ii\phi_{C_2}}$ (and $D_2$ analogously), and then yields
\begin{multline}
(\kappa-\varphi)^2 \left\| \begin{pmatrix} \Re(C_2+D_2) & \Im(C_2+D_2) \\ \Im(-C_2+D_2)& \Re(C_2-D_2) \end{pmatrix} \begin{pmatrix}\cos \frac{\phi_{C_2}+\phi_{D_2}}2 \\ \sin \frac{\phi_{C_2}+\phi_{D_2}}2\end{pmatrix}\right\|^2\\
 = (\kappa-\varphi)^2\left( |C_2|+|D_2|\right)^2.
\end{multline}

The $Z$-component can be rewritten as
\begin{multline}
\left[ \cos \alpha \Re \left(\delta (C_2^*+D_2)\right) +\sin \alpha \Im \left(\delta^* (C_2-D^*_2)\right) \right]^2\\
 = \left| \delta\right|^2 \left[ \Re \left( |C_2| \ee{\ii (\alpha+\arg \delta-\phi_{C_2})} +|D_2| \ee{\ii( \alpha-\arg \delta-\phi_{D_2})} \right)\right]^2.
\end{multline}
where we used $\delta=|\delta| \ee{\ii\arg\delta}$. To maximize this,  the arguments of the imaginary exponents need to vanish. This is achieved by
\begin{eqnarray}
\alpha=\frac{\phi_{C_2}+\phi_{D_2}}2, \qquad \arg \delta=\frac{\phi_{C_2}-\phi_{D_2}}2,
\end{eqnarray}
which is the same condition for $\alpha$ as derived before.
(Alternatively, here and also for the $X-Y$-component, we obtain a second solution by replacing $\alpha\to \alpha+\pi$ and $\arg \delta\to \arg \delta+\pi$. However, this just corresponds to multiplying $\brho_1$ by $-1$.)

This shows that Alice's optimal choice of initial states is independent of Bob's initial state and given by \eqref{eq:aliceoptstate}. In order to ensure a maximal leading order signal strength, as quantified by the trace distance between output states, Bob needs to initialize his detector in a pure state with off-diagonal density matrix element $\delta=|\delta|\ee{\ii({\phi_{C_2}-\phi_{D_2}})/2}$. The Bloch vector of such an optimal initial state for Bob is then
\begin{eqnarray}
\brho_{B,0}=\begin{pmatrix} 2|\delta| \cos\frac{\phi_{C_2}-\phi_{D_2}}2\\-2|\delta| \sin\frac{\phi_{C_2}-\phi_{D_2}}2 \\ \varphi-\kappa
\end{pmatrix}.
\end{eqnarray}
This is a pure state of the form $\sqrt{\varphi}\ket{e_B}+\sqrt{\kappa}\ee{ - \ii({\phi_{C_2}-\phi_{D_2}})/2}$. With these optimal choices of $\alpha$ and $\arg \delta$ we obtain
\begin{eqnarray}
\left\| \bi{M}_2 \brho_i\right\|^2 = \left( (\kappa-\varphi)^2+ 4 |\delta|^2\right) \left(|C_2|+|D_2|\right)^2=\left(|C_2|+|D_2|\right)^2
\end{eqnarray}
where we used that $|\delta|=\sqrt{ \kappa\varphi}$ for  pure initial states of Bob.

Therefore, if Bob initializes his detector in an optimal state, then the trace distance between the final states resulting from Alice's optimal input states is
\begin{eqnarray}
D\left(\xi(\brho_1),\xi(\brho_2)\right) \sim |C_2|+|D_2|+\mathcal{O}(\lambda^4).
\end{eqnarray}

It is interesting to note that Bob, in order to distinguish the two final states of his detector, has to perform the measurement on his detector state with respect to a basis that is mutually unbiased with respect to a basis containing his initial state. Because the final states of his detector have Bloch vectors
\begin{eqnarray}
\bi{M}_2 \brho_{1,2} = \pm\left(|C_2|+|D_2|\right)\begin{pmatrix} (\kappa-\varphi) \cos\frac{\phi_{C_2}-\phi_{D_2}}2 \\ -(\kappa-\varphi)\sin\frac{\phi_{C_2}-\phi_{D_2}}2 \\2| \delta|\end{pmatrix}
\end{eqnarray}
which are orthogonal to the Bloch vector $\brho_{B,0}$ of his initial state.

\section{Classical Channel Capacity from Leading Order Signaling}\label{sec:capacity}
In this section we review which classical channel capacity arises from the optimal use of leading order signals that we found in the previous section. 
We consider different measures for the classical capacity: First, the probability for the successful transmission of one bit in a single use of the channel, and then capacities arising from the repeated use as a classical channel and the Holevo capacity of the channel. 

We find that the perturbative expansion of all these capacity measures is, to leading order, determined by (different powers of) $|C_2|+|D_2|$ which in the previous section we found to be the optimal leading order trace distance between Bob's two possible output states. This suggests that $|C_2|+|D_2|$ is well suited as a general measure for the signal strength between two particle detectors, within the regime of perturbative interactions.


\subsection{Success Probability of Transmitting One Bit in a Single Use}
This first measure for the channel's classical capacity can be motivated by a communication task or game where Alice needs to transmit a random bit to Bob: First, Alice is given a random bit, which with equal probability $1/2$ is `0' or `1'.
To communicate the bit to Bob, Alice may use the quantum channel one time. In the end, for Alice and Bob to win the game, Bob has to tell the correct value of the bit which was given to Alice. When can Alice and Bob win the game with a probability higher than $1/2$, the success probability when Bob just makes a random guess?

If Alice is able to alter the probability of Bob to find a particular measurement outcome from one probability $p$ to another value $q\neq p$, then Bob and Alice can use this influence to win the game with a probability of \cite{jonsson_information_2015,jonsson_decoupling_2016}
\begin{eqnarray}
P_{\text{bit}}=\frac12+\frac{|p-q|}2.
\end{eqnarray}
This means that any influence of Alice on Bob allows them to improve the success probability above $1/2$.
By using the optimal initial states and measurements identified in the previous section, Alice and Bob can maximize the difference for the measurement outcome to
\begin{eqnarray}
P_{\text{bit}}=\frac12+\frac12 D(\xi(\rho_1),\xi(\rho_2))=\frac12+|C_2|+|D_2|+\mathcal{O}(\lambda^4).
\end{eqnarray}

The success probability for the transmission of one bit in a single channel has been used as a measure of signal strength in \cite{jonsson_information_2015,blasco_violation_2015,blasco_timelike_2016,simidzija_information_2017}. There, however,  Alice's and Bob's initial states were not optimized. The optimal achievable leading order contribution presented here may thus be able to increase the  estimates  in these  works.

\subsection{Repeated Use as Classical Channel}
Single-shot capacity measures, which only consider a single use of the channel, are, arguably, best suited to analyze the channel between two particle detectors communicating via a quantum field. This is because measures like the Shannon capacity or the Holevo capacity are defined as asymptotically achievable transmission rates in the limit of many identical channel uses. 

Having Alice and Bob repeatedly communicate via the quantum field does not constitute many uses of an identical channel. Instead, due to timelike signaling propagation and due to the vacuum entanglement of the field, there may arise correlations between the different interactions of Alice and Bob with the field. 
Therefore, the quantum channel map arising from Alice's and Bob's first coupling to the field is not strictly identical to the channel arising from their second, or any subsequent coupling to the field. However, allowing for enough time to pass between the different couplings should, generally speaking, decrease these correlations far enough to become irrelevant. A detailed investigation of this question should be an interesting question for future research.

Whereas  the applicability of asymptotic capacity measures to the channel at hand may be limited, it is still interesting to evaluate them if only for comparisons with other qubit channels.
As an example of two of such measures, we review results on the leading order contributions to the Shannon capacity in the following, and to the Holevo capacity  in the subsequent section.

The Shannon capacity that arises from a repeated use of the channel as a classical, binary asymmetric channel was already considered in  \cite{jonsson_information_2015}. Here Alice and Bob are allowed to use the channel for a large number of times, however Bob has to perform separate measurements on each of the individual channel outputs. (In contrast to the Holevo capacity which we consider below.) Therefore, each channel use corresponds to the use of a classical, binary asymmetric channel \cite{silverman_binary_1955}. The Shannon capacity then gives the information measured in bits per channel use that can be reliably transmitted from Alice to Bob in the limit of large numbers of channel uses \cite{nielsen_quantum_2010}. Using known results on the binary asymmetric channel \cite{silverman_binary_1955}, a perturbative expansion for the Shannon capacity for the channel between particle detectors was given in \cite{jonsson_information_2015} which, however, did not yet  include the optimization of the detectors' initial states. Using the optimal initial states  yields the expansion
\begin{eqnarray}
C_{\text{Shannon}}\sim \frac2{\ln 2} \left( |C_2|+|D_2|\right)^2+\mathcal{O}(\lambda^6).
\end{eqnarray}
for the Shannon capacity which, to leading order in perturbation theory, is proportional to $(|C_2|+|D_2|)^2$. 

\subsection{Holevo Capacity}
The rate of classical information transmission can be improved upon by allowing Bob to perform joint measurements on all the outputs obtained from multiple, parallel channel uses. (Alice is still required to prepare separable states over the different channel inputs.) 
The rate of bits per channel use that can be reliable transmitted under these conditions is captured by the Holevo capacity of the channel. Denoting the channel by$\xi$ it reads \cite{nielsen_quantum_2010},
\begin{eqnarray}
C_{\text{Holevo}}= \max_{p_j,\rho_j} S\left(\xi\left(\sum_j p_j\rho_j\right)\right) -\sum_j p_j S\left(\xi\left(\rho_j\right)\right)
\end{eqnarray}
where the choice of Alice's ensemble of initial states $\rho_i$ and their relative frequencies $p_i$ need to be optimized.

Applying results from \cite{berry_qubit_2005}, it is possible to show that the Holevo capacity is maximized by an ensemble of signaling states containing only two states, and that the Holevo capacity can be expanded as \cite{jonsson_decoupling_2016}
\begin{eqnarray}\label{eq:holevocap}
C_{\text{Holevo}}\sim -\ln(P_2) \frac{(|C_2|+|D_2|)^2}{4\ln2}+\mathcal{O}(\lambda^6),
\end{eqnarray}
where it is assumed that Bob initializes his detector in the ground state. This result on the Holevo capacity of the channel completes early results in \cite{cliche_relativistic_2010} which are restricted to energy eigenstates as initial states of Alice.

In the leading order of the Holevo capacity  the channel coefficient $P_2$ appears, besides the quantity $|C_2|+|D_2|$.
This coefficient is the leading order contribution to the probability of Bob's detector to get excited, from its ground  to its excited state, by the interaction with the field vacuum. It is independent of Alice's presence.

Since the coefficient is of order $P_2\sim\mathcal{O}(\lambda^2)$ itself, its first effect is that the leading order contribution of the Holevo capacity is of higher order than the previous Shannon capacity, which is only of order $(|C_2|+|D_2|)^2\sim\mathcal{O}(\lambda^4)$.

On the other hand, it makes the Holevo capacity sensitive to the effects that arise from Bob's interaction with the field alone, which are captured also in the channel coefficients $Q,R,S$ besides $P$ in \eqref{eq:chanstructgeneral}, and could be considered as noise contributions to the channel.
On first sight it may even look as if when  $P_2$ decreases, arbitrary high leading order contributions to the Holevo capacity can be achieved. However, decreasing $P_2$ typically requires the use of switching functions that only change slowly. These, in turn, tend to decrease the value of $|C_2|+|D_2|$ which  would counter the increase of the $\ln (P_2)$ factor.

\section{Relativistic Motion in Minkowski Spacetime}\label{sec:minkowski}

In this section we discuss how inertial and accelerated motion impacts signaling between particle detectors in Minkowski spacetime. We find that the leading order signal strength is dominated by the classically expected relativistic effects. To maximize the signal strength between inertially moving detectors, the detectors have to be detuned such as to correct for the relativistic Doppler effect. When either party is uniformly accelerated, the acceleration horizon and the infinite Doppler shift at early and late times limit the leading order signal strength.

In particular, we will here study signaling in a massless Klein-Gordon field. (The previous sections made no assumptions on the mass of the field.) The propagation behaviour of signals in massless fields in (3+1)-dimensional Minkowski spacetime is generally familiar. It is captured by the field commutator  
\begin{eqnarray}\label{eq:comm31D}
\fl\qquad\comm{\phi(\bi{x},t)}{\phi(\bi{x}',t')}=\frac\ii{4\pi} \frac1{\left|\bi{x}-\bi{x}'\right|} \left( \delta\left(t'-t-\left|\bi{x}-\bi{x}'\right|\right) -\delta\left( t'-t+\left|\bi{x}-\bi{x}'\right|\right)\right)
\end{eqnarray}
which has support only on the boundary of the lighcone. Therefore, in (3+1) dimensions signals propagate strictly at the speed of light.

However, in (1+1)-dimensional and in (2+1)-dimensional Minkowski spacetime  signaling between timelike separated spacetime points is possible \cite{jonsson_information_2015,jonsson_information_2016}. This is the case because the commutator in (1+1)-dimensional Minkowski spacetime, 
\begin{eqnarray}\label{eq:comm11D}
\fl\qquad\comm{\phi({x},t)}{\phi({x}',t')}=\frac\ii2 \text{sgn}(t'-t) \theta\left( (t-t')^2 - (x-x')^2\right),
\end{eqnarray}
and in  (2+1)-dimensional Minkowski spacetime,
\begin{eqnarray}\label{eq:comm21D}
\fl\qquad\comm{\phi(\bi{x},t)}{\phi(\bi{x}',t')}=\frac\ii{4\pi^2} \frac{ \text{sgn}(t'-t)}{\sqrt{(t'-t)^2-\left|\bi{x}-\bi{x}'\right|^2}} \theta\left((t'-t)^2-\left|\bi{x}-\bi{x}'\right|^2\right),
\end{eqnarray}
have support inside the lightcone. (Detailed calculations of the commutator in Minkowski spacetime are found, e.g., in \cite{martin-martinez_causality_2015,jonsson_decoupling_2016}.)

We show that there  is a characteristic difference between null and timelike signals: For null signals it is important that the detectors are resonant in order to maximize the signal strength. In fact, if receiver and sender are resonant  the leading order signal strenght can always be increased by increasing the interaction time for which the detectors couple to the field. In contrast, the leading order signal strength of timelike signals is bounded, and sender and receiver do not need to synchronize their detectors either. Instead, they only need to individually optimize their switching times with respect to their own detector energy gaps.

We also show that in general the transmission of information via the amplitude of a massless field does not rely on the transmission of a minimum amount of energy from the sender to the receiver. Because the leading order signal strength decays slower than the energy density of the signal when the distance between sender and receiver is increased. We show that this property, which was shown for timelike signals in \cite{jonsson_information_2015}, is also shared by null signals in all dimensions.

\subsection{Simplification and Time-mirror symmetry of Leading Order Signaling}

In this section we discuss some properties of the integral terms of the channel coefficients $C_2$ and $D_2$ which simplify the subsequent evaluation of the signal strength. We show that the integration boundaries in the original definition of the coefficients can be simplified, that the signal strength is preserved under time-inversion, and that the use of sudden switching function does not introduce divergences.

To begin with we note that a solution of $D_2$, as defined in \eqref{eq:d2}, can always be obtained from a solution for $C_2$, as defined in \eqref{eq:c2}, from $D_2(\Omega_A,\Omega_B)=-C_2(\Omega_A,-\Omega_B)$, i.e., by changing the overall sign and $\Omega_B\to-\Omega_B$ the sign of Bob's detector energy gap. Therefore, in the following, we focus our discussion on $C_2$ only.

In the integral in $C_2$ in \eqref{eq:c2}, the outer (coordinate) time variable $t_1$ sets the upper integration boundary for the inner time variable $t_1$. This dependency arises directly from the Dyson series expansion used for the time evolution operator (see \eqref{eq:dysonseries}).
However, the boundary for the $t_2$-integral is actually always lower because the commutator of the field vanishes at spacelike separations. We furthermore assume that Alice couples to the field only during the finite time window $0\leq t_2\leq T_A$, and Bob only during $T_1\leq t_1\leq T_2$.
Therefore, we can tighten the integration boundaries to
\begin{eqnarray}\label{eq:C2tightbounds}
\fl C_2= \lambda_A \lambda_B\integral{t_1}{T_1}{T_2}\integral{t_2}{0}{\min\left(T_A,\tilde{t}(t_1),t_1\right)} \chi_A(t_2) \chi_B(t_1) \ee{\ii (\Omega_B\tau_B(t_1)- \Omega_A\tau_A(t_2) )} \comm{\phi(x_A(t_2)}{\phi(x_B(t_1))},\nonumber\\*
\end{eqnarray}
where $\tilde{t}(t_1)$ is the coordinate time at which the lightray that reaches Bob at coordinate time $t_1$ emanated from Alice. 

In particular, the integration boundaries of $t_1$ and $t_2$ are interdependent, only if the Alice and Bob are in null contact. When Bob is located inside the future lightcone of Alice the $t_2$ integral always is performed over $0\leq t_2\leq T_A$. Below we will see that this is the reason why for lighlike signals the resonance between Alice's and Bob's detector frequencies increase the signal, whereas it is irrelevant for timelike signals.

The integral expressions for $C_2$ and $D_2$ have another property which halves the number of scenarios with moving detectors that we need to investigate, which is a time-mirror symmetry: The leading order signal strength $|C_2|+|D_2|$ of a given scenario is the same as in the scenario we obtain by ``running the movie backwards'', i.e., when time is inverted such that the detectors move backwards and Bob now is the sender instead of being the receiver. For example, this means, that the signal strength from a resting sender to a receiver accelerating away, is the same as from a sender with opposite acceleration to a resting receiver.

To show this we assume, without loss of generality, that $\tau_A(t=0)=\tau_B(t=0)=0$. The wordlines of the time mirrored-scenario are then given by $\bi{x}_d'(t)=\bi{x}_d(-t)$ such that the detector proper times are $\tau_d'(t)=-\tau_d(-t)$, and the mirrored switching functions are $\chi_d'(t)=\chi_d(-t)$. Using this, it is straightforward to show that the leading order signaling coefficients in the mirrored scenario are \cite{jonsson_decoupling_2016}
\begin{eqnarray}
C_2'= C_2, \qquad D_2'= -D_2^*,
\end{eqnarray}
such that the optimal signal strength is the same for the original and the mirrored scenario
\begin{eqnarray}
|C_2'|+|D_2'|=|C_2|+|D_2|.
\end{eqnarray}

The signaling coefficients $C_2$ and $D_2$ are less prone to divergences than the coefficients $P_2,Q_2,R_2,S_2$, which describe the local effects of Bob's detector interacting with the vacuum of the field alone. This allows us to use sudden switching functions in our study of the leading order signal strength, i.e., functions that abruptly jump from 0 to 1 inside the time interval during which the  detector is interacting with the field.

In the single-detector coefficients such sudden switching functions lead to UV-divergences. Instead the coupling needs to be switched through smooth coupling functions \cite{satz_then_2007,hodgkinson_how_2012}. Also, the signaling coefficients are not affected by the IR-divergence arising from the zero mode of the massless field in (1+1)-dimensional Minkowski spacetime \cite{martin-martinez_particle_2014}.

\subsection{Detectors at Rest}\label{sec:rest} 
In this section we show that the signal strength of null signals decays slower than the energy content of the signal when the distance between sender and receiver is increased. This was already shown to be the case for timelike signals in \cite{jonsson_information_2015}.
To this end we study the leading order signal strength in the simple scenario where Alice and Bob both are at rest in Minkowski spacetime at a fixed distance $L$ from each other.

We want to compare Minkowski spacetime from 1+1 to 3+1 dimensions. 
Since in 3+1 dimensions signals propagate only between null separated points we choose Bob's coupling to be  null separated from Alice's coupling. (We discuss the timelike signals which appear in lower dimensions later, in Section \ref{sec:timelike}.)
This means that Bob couples to the field only in the time interval during which the lightrays reach Bob which emanate from Alice while she is coupling to the field. So  Alice couples to the field for $0\leq t_1\leq T_A$, and Bob for $L\leq t_2\leq T_A+L$. 

\begin{figure}
\centering
\includegraphics[trim={0 2.5cm 2cm 1.5cm},clip,width=0.6\columnwidth]{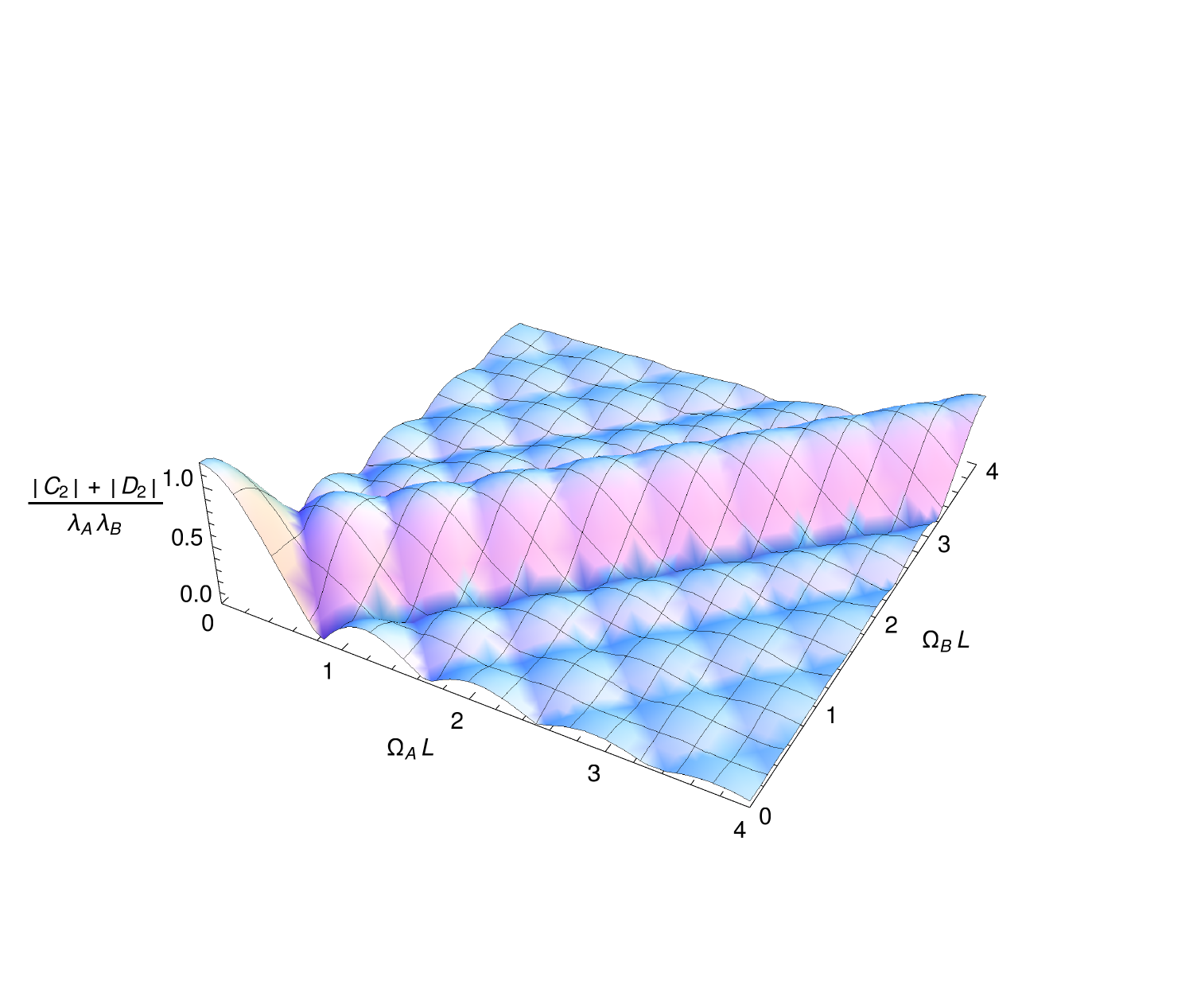}
\caption[Resonance of signal strength for resting detectors]{Leading order signal strength $|C_2|+|D_2|$ (in multiples of $\lambda_A\lambda_B$) from \eqref{eq:C2in3+1Drest}, for two detectors at rest in 3+1D Minkowski space at distance $L$ from each other, with energy gaps $\Omega_A$ and $\Omega_B$. The signal strength is maximal  for resonant detectors, along the diagonal $ \Omega_A= \Omega_B$. The peak becomes more distinct for larger interaction times $T_A$. Here,  the interaction time is $T_A=7.5 L$. (Figure adapted from \cite{jonsson_decoupling_2016}).
}
\label{fig:resonanceC2D2}
\end{figure}

With these coupling times there are analytical solutions to the integrals of the leading order signaling contributions \cite{jonsson_decoupling_2016}. In 1+1 dimensions we obtain
\begin{eqnarray}
\fl \quad C_2&= \lambda_A \lambda_B \frac{\ii  \ee{\ii \Omega_B L}}{2 \Omega_A \Omega_B ( \Omega_A- \Omega_B)} \left( ( \Omega_B- \Omega_A) \left(1-\ee{\ii \Omega_B T_A}\right) + \Omega_B \left( \ee{\ii (\Omega_B- \Omega_A) T_A} -1\right)\right),
\end{eqnarray}
and in 3+1 dimensions we obtain
\begin{eqnarray}\label{eq:C2in3+1Drest}
C_2&= \lambda_A \lambda_B \frac{\ee{\ii \Omega_B L} \left(1-\ee{\ii  ( \Omega_B- \Omega_A) T_A}\right) }{4 \pi L ( \Omega_A- \Omega_B)}.
\end{eqnarray}
As shown in Figure \ref{fig:resonanceC2D2}, the signal strength is maximal for resonant detectors with equal detector gaps $\Omega_A=\Omega_B$. The reason for this is that one of the terms resulting from the upper integration boundary $t_2\leq t_1-L$ becomes non-oscillatory when $\Omega_A=\Omega_B=\Omega$.
With such identical detectors the leading order signaling coefficients are  in 1+1 dimensions
\begin{eqnarray}
\fl \quad C_2=-\lambda_A \lambda_B\frac{\ee{\ii \Omega L} \left(\ii \left( \ee{\ii \Omega T_A}-1\right) + \Omega T_A\right)}{2 \Omega^2}, \quad  D_2=\lambda_A \lambda_B\frac{\ii\ee{-\ii (\Omega L + 2 \Omega T_A)} \left( \ee{\ii  \Omega T_A}-1\right)^2}{4 \Omega^2},
\end{eqnarray}
and in 3+1 dimensions 
\begin{eqnarray}\label{eq:restC231D}
\fl\quad C_2=\lambda_A \lambda_B\frac{\ii \ee{\ii \Omega L} T_A}{4 \pi L},\qquad  D_2= -\lambda_A \lambda_B\frac{\ee{-\ii \Omega (L+2 T_A)} \left( \ee{\ii 2 \Omega T_A}-1\right)}{8\pi \Omega L}.
\end{eqnarray}

The leading order signal strength is maximized in the limit of vanishing detector frequency $\Omega\to0$, i.e. for a zero-gap detector. The detector Hamiltonian of such a detector vanishes (or  is proportional to the identity operator). This means that the detector has no distinguished energy eigenstates and  ceases to have a free time evolution of its own. The latter is also the reason why zero-gap detectors maximize the signal strength.
With the rotation from a non-zero energy gap in place, Alice's initial state is rotated around the $Z$-axis of the Bloch sphere while coupling to the field. This diminishes the signal strength because it leads to an averaging effect over each detector period. However, when the detector gap is zero, Alice can prepare an eigenstate of the interaction Hamiltonian in her detector which remains unchanged and, thus, yields a stronger signal.
Zero-gap detectors are also interesting because they allow for a comparison of the perturbative analysis to the non-perturbative solutions of  \cite{landulfo_nonperturbative_2016}.

If Alice and Bob both use a zero-gap detector the signal strength is in 1+1 dimensions
\begin{eqnarray}
 \left|C_2\right|+|D_2|= \lambda_A \lambda_B \frac{T_A^2}{2},
\end{eqnarray}
and in 3+1 dimensions
\begin{eqnarray}
 |C_2|+|D_2|= \lambda_A \lambda_B \frac{T_A}{2\pi L}.
\end{eqnarray}
For zero-gap detectors there is also an analytic solution in 2+1 dimensions:\footnote{We note that the corresponding formula in \cite{jonsson_decoupling_2016} contains a typo. It should, as in \eqref{eq:21Danalyticalatrest}, contain the difference of the logarithm and the square root term, but not their sum.}
\begin{eqnarray}\label{eq:21Danalyticalatrest}
\fl \quad |C_2|+|D_2|= \frac{\lambda_A \lambda_B}\pi \left( (T_A+L) \ln \left(1+\frac{T_A+\sqrt{2L T_A+T_A^2}}L \right) -\sqrt{2L T_A+T_A^2} \right).
\end{eqnarray}

The scenario of two detectors at rest allows us to analyze how the signal strength depends on the distance between Alice and Bob. In 1+1 dimensions the signal strength is independent of the distance: This is what one would expect because the surface of the propagating wave front does not expand since there is only a single spacelike dimension.

In higher dimensions one might intuitively expect that the signal strength would decay proportional to the expanding surface of the propagating wavefront, i.e., as $\sim 1/L$ in 2+1 dimensions, and as $\sim 1/L^2$ in 3+1 dimensions. This is the rate at which the energy density of the signal has to decay as the total energy of the signal is dispersed over the increasing spherical wavefront surface.

However, we find that the signal strength only decays as $\sim 1/L$ in 3+1 dimensions, and is $\sim 1/\sqrt{L}$ in 2+1 dimensions (both in the analytical solution for zero-gap detectors above, as well as in numerical solutions for gapped detectors). In fact, this behaviour is to be anticipated already from the dependency of the field commutator on the spatial distance between the field operators.
The ratio of signal strength to energy density of the signal thus grows unbounded as $L$ is increased. 

This shows that the flow of information carried by the amplitude of a massless field should in general not be thought of as being tied to a certain, minimum flow of energy, both for timelike and for null separations between sender and receiver.
In order to store information into the field the sender always has to invest a certain amount of energy, as dictated, e.g., by the results of \cite{caves_quantum_1994}. However, as we observe here, the propagation of information can decouple from the propagation of energy inside the massless field.
The most distinct occurrence of this phenomenon is  represented by timelike signals in 1+1 dimensions which carry information without carrying any energy from the sender to the receiver \cite{jonsson_information_2015,jonsson_information_2016}.

\subsection{Detectors in Inertial Motion}\label{sec:inertial}
Signaling between two detectors which are moving inertially with respect to each other in Minkowski spacetime, is the first scenario in which we observe the impact of a relativistic effect. When the detectors are in inertial motion the leading order signal strength is not maximal for detectors with identical energy gaps anymore, which it was for detectors at rest. Instead the detectors have to be detuned so as to account for the relativistic Doppler effect.

If the detector energy gaps are tuned to correct for the Doppler shift between sender and receiver, then the leading order signal strength grows with longer interaction times $T_A$, as for detectors at rest. This resonance effect arises from the same mechanism as above due to the nested integral structure. When the detector frequencies are optimally chosen one of the terms arising from the upper integration boundary of the inner $t_2$-integration becomes non-oscillatory.

To illustrate this, we look at a scenario where Alice and Bob move apart from each other. We choose Bob's rest frame as coordinate system, i.e., Bob remains at rest at $\bi{x}_B=0$. Alice is moving away from Bob at constant speed $0\leq v<1$. She couples to the field for her proper time interval $0\leq \tau_A\leq T_A$. Without loss of generality, we choose $t(\tau_A=0)=0$ and we denote the distance between Alice and Bob at time $t=0$ by $L$.

As above, we restrict Bob to only couple to the field during the time interval for which he  is null separated from Alice. The first of Alice's lightrays reaches Bob at $t=L$. The last of Alice's lightray, taking into account Alice's motion and relativistic time dilation, reaches Bob at $t= T_A (1+v)/\sqrt{1-v^2}=\zeta T_A$, where 
\begin{eqnarray}
\zeta=\sqrt{\frac{1+v}{1-v}}
\end{eqnarray}
is the relativistic Doppler factor. Therefore, we couple Bob to the field during $L\leq t\leq L+ \zeta T_A$.

\begin{figure}
\centering
\includegraphics[width=0.6\columnwidth]{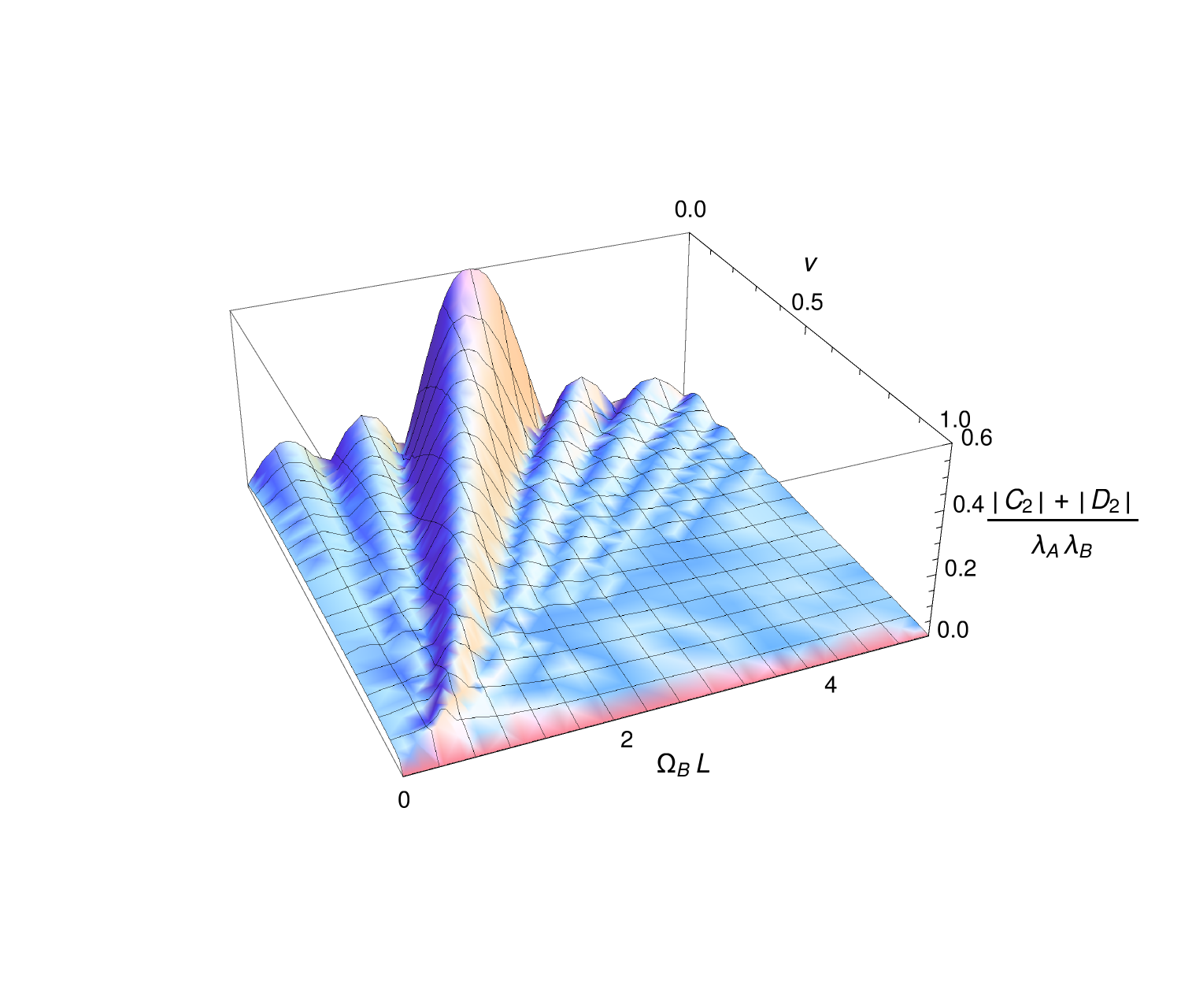}
\caption[Doppler shift for moving detectors]{Leading order signal strength $|C_2|+|D_2|$ (in multiples of $\lambda_A\lambda_B$) between two detectors moving apart inertially with speed $v$. The relativistic Doppler effect requires Bob to detune his detector energy gap to $\Omega_B=\Omega_A/ \zeta$ in order to correct for the red-shift of Alice's detector gap which is set to $\Omega_A=2.5$. The initial distance between the detectors is $L$. Alice is coupled to the field for a proper time of $T_A=7.5 L$.  (Figure adapted from \cite{jonsson_decoupling_2016}.}
\label{fig:doppler}
\end{figure}

In 3+1 dimensions this results in \cite{jonsson_decoupling_2016}
\begin{eqnarray}
\fl C_2
&=\lambda_A \lambda_B\frac{\ii\sqrt{1-v^2}}{4\pi v} \ee{\ii L \left( \Omega_B +\frac{\sqrt{1-v^2}}{v}( \Omega_A- \zeta \Omega_B)\right)}\nonumber\\*
\fl&\quad\times \left( \Gamma\left(0,\frac{\ii L \sqrt{1-v^2}}v ( \Omega_A- \zeta \Omega_B)\right) -\Gamma\left(0,\frac{\ii (T_A v+L\sqrt{1-v^2})}v ( \Omega_A- \zeta \Omega_B)\right) \right)
\end{eqnarray}
with the incomplete Gamma function $\Gamma(a,x)=\integral{t}{x}{\infty} t^{a-1}\ee{-t}$.
Figure \ref{fig:doppler} shows that the signal strength is maximal when Bob exactly accounts for the Doppler red-shift of Alice's detector that lowers his detector energy gap to $\Omega_B= \Omega_A/ \zeta$. Then we obtain
\begin{eqnarray}
C_2=\lambda_A \lambda_B\frac{\ii\ee{\ii \Omega_B L}}{4\pi} \frac{\sqrt{1-v^2}}v \ln\left(1+\frac{v T_A}{L \sqrt{1-v^2}}\right).
\end{eqnarray}
This correctly reproduces the result \eqref{eq:restC231D}  for resting detectors in the limit $v\to0$. However, for $v>0$, the signal strength for resonant moving detectors remains lower than for resonant resting detectors because the field commutator \eqref{eq:comm31D} decays as the distance between the detector increases with time.

Also in (1+1)-dimensional Minkowski spacetime the leading order signaling coefficients are analytically solvable \cite{jonsson_decoupling_2016},
\begin{eqnarray}
C_2&=-\lambda_A \lambda_B\frac{\ee{\ii \Omega_B L}}{2 \Omega_A} \left( R+\frac\ii{ \Omega_B}\left(\ee{\ii \Omega_B \zeta T_A}-1\right)\right),
\end{eqnarray}
where
\begin{eqnarray}
R&=\begin{cases} \zeta T_A &\text{ if } \Omega_B= \Omega_A/ \zeta \\ \frac{2\ii}{\Omega_B-\Omega_A/\zeta} \ee{\ii ( \Omega_B \zeta - \Omega_A) \frac{T_A}2} \sin\left(( \Omega_B \zeta - \Omega_A) \frac{T_A}2\right) &\text{ else} \end{cases}.
\end{eqnarray}

We have considered a scenario where Alice and Bob move apart and have to correct for the red-shift of their detector energy gaps. 
It is interesting to note that, due to the time-mirror symmetry of the signal strenght, this analysis also covers the scenario of Alice and Bob moving towards each other, despite that there is a blue-shift occuring when Alice and Bob move towards each other.
In the time-mirrored scenario Bob is the sender. Therefore, the lowering of his detector energy gap $\Omega_B=\Omega_A/\zeta$ with respect to Alice's gap, chosen to account for the red-shift in the original scenario, now exactly corrects for the blue-shift when moving towards Alice.

\subsection{Signaling Across an Acceleration Horizon}\label{sec:accelerated}
When Alice and Bob are in inertial motion, as in the previous sections, the integrals appearing in the  leading order signal contribution can grow arbitrarily large if the detectors couple to the field for an unlimited  time.
This ceases to be the case if either Alice or Bob are in uniformly accelerated motion. Here, the  leading order signal strength across the acceleration horizon is bounded  even if Alice and Bob can couple to the field for an infinite time.
We show this by calculating analytical solutions to the leading order signaling contributions in (1+1)-dimensional and (3+1)-dimensional Minkowski space.

Studying quantum  communication with uniformly accelerated observer is of particular interest because they experience the Unruh effect \cite{unruh_notes_1976}. And the question has therefore also been addressed in earlier works such as \cite{bradler_quantum_2012,downes_quantum_2013}.
The Unruh effect impacts the quantum channel from Alice to Bob through the coefficients $P,Q,R,S$ that appear in Bob's final state \eqref{eq:chanstructgeneral}. 
In contrast, the effects that limit   the leading order signal strength  $|C_2|+|D_2|$ arise from the classical relativistic Doppler shift of the accelerated party which is infinitely large at early and late times.

The worldline of a uniformly accelerated observer in $n$+1 dimensional Minkowski spacetime can be parametrized as
\begin{eqnarray}
t(\tau)=\frac1a \sinh(a\tau), \qquad x^1=\frac1a \cosh(a\tau)
\end{eqnarray}
and $x^2=...=x^n=0$, where $\tau$ is the observer's propertime. In this choice of coordinates the accelerated observer can receive signals from the past lighcone of the origin of the coordinate system at $x^0=x^1=...=x^n=0$, but cannot send any signals there. Conversely, the accelerated observer can send signals into the future lightcone of $x^\mu=0$ but cannot receive any signals from there.

We here study the leading order signal strength between an accelerated observer and an observer that sits right behind the acceleration horizon. This means that signals can only travel across the acceleration horizon from the sender to the receiver but not back in the opposite direction, i.e.,  the sender influences the receiver's final state, but the receiver's presence has no influence on the sender's final state.

In particular, we will put Alice, the sender, on the accelerated worldline above, and put Bob at rest at the origin of the coordinate system. Alice couples to the field along her entire worldline, whereas Bob only couples to the field for $t>0$.
This means we set the switching function of Alice in \eqref{eq:C2tightbounds} constantly to $\eta_A(\tau_A)=1$, whereas Bob, whose proper time coincides with the coordinate time, is switched by the Heaviside function $\chi_B(t)=\theta(t)$.

Due to the mirror symmetry of the leading order signal strength, the signal strength of this scenario  is the same as in the mirrored scenario. There Alice would be at rest at the origin and couple to the field for $t<0$ whereas Bob would be accelerated and coupling to the field all along his worldline.

\begin{figure}[tb]
\centering
\subfloat[ 
\label{fig:3dplot}]{
  \includegraphics[width=0.45\columnwidth]{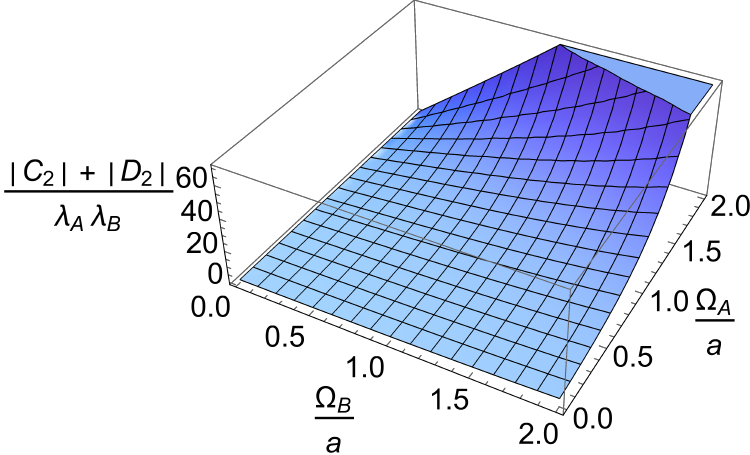}
  }  
\subfloat[ 
  \label{fig:detail}]{
    \includegraphics[width=0.45\columnwidth]{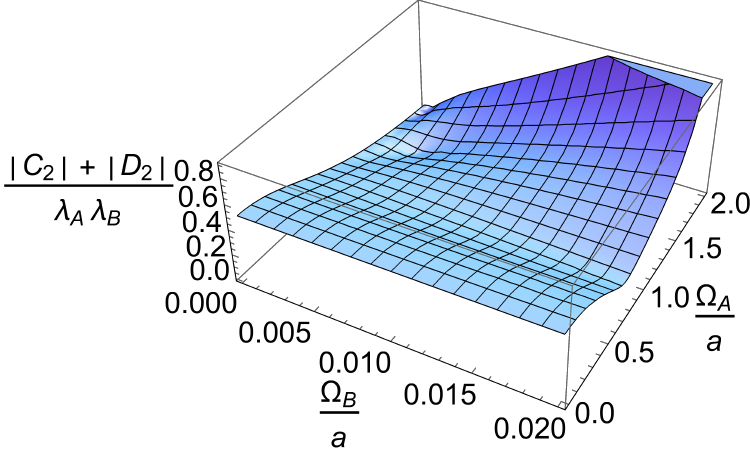}
    }
\caption[Signal strength across acceleration horizon]{
The leading order signal strength $|C_2|+|D_2|$ 
across the acceleration horizon in 3+1D Minkowski spacetime is a function of the ratios $\omega_A/a$ and $\Omega_B/a$ between the detector energy gaps and the sender's proper acceleration $a$. 
Fig.~\ref{fig:detail} shows non-trivial behaviour of the total signal strength for low $\Omega_B/a$ not resolved in Fig.~\ref{fig:3dplot}.
}
\label{fig:accelerated31D}
\end{figure}

In (3+1)-dimensional Minkowski spacetime the leading order signal strength depends only on the ratios $x=\Omega_B /a$ and $y=\Omega_A/a$ between the detector energy gaps and Alice's proper acceleration. 
This is because the coupling constant $\lambda$ is dimensionless, such that the detector gaps and the acceleration are the only physical scales that enter the problem. Figure \ref{fig:accelerated31D} shows a plot of this leading order signal strength $|C_2|+|D_2|$.
It can be obtained by using the sender's proper time as integration variable in \eqref{eq:c2}\footnote{Note that earlier versions of this article here cited an erroneous expression from \cite{jonsson_decoupling_2016}.}. 
\begin{multline}\label{eq:C2corr}
C_2 =\lambda_A \lambda_B\integral{t_1}{}{}\integral{t_2}{}{t_1} \chi_A(t_2) \chi_B(t_1) \ee{\ii (\Omega_B\tau_B(t_1)- \Omega_A\tau_A(t_2) )} \comm{\phi(x_A(t_2))}{\phi(x_B(t_1))}\\
=\lambda_A \lambda_B\integral{t_1}{0}{\infty}\integral{\tau_A}{-\infty}{\infty}   \ee{\ii (\Omega_B t_1- \Omega_A\tau_A )} \frac\ii{4\pi}\frac{a}{\cosh(a\tau_A)}\delta\left( t_1-\frac1a \ee{a\tau_A} \right)  \\
=\frac{\ii\lambda_A \lambda_B}{4\pi} \integral{t_1}{0}{\infty}\integral{\tau_A}{-\infty}{\infty}   \ee{\ii (\Omega_B t_1- \Omega_A\tau_A )} \frac{2a^2t_1}{a^2(t_1)^2+1}\frac1{\left|a t_1\right|}\delta\left( \tau_A-\frac1a\ln\left(a t_1\right) \right) \\
=   \frac{-\ii \lambda_A \lambda_B \,\mathrm{csch}(\pi  y)}2 \left(\frac{e^{\frac{\pi  y}{2}} \left(\frac{1}{x}\right)^{-1-i y} \, _1F_2\left(1;\frac{i y}{2}+1,\frac{i
   y}{2}+\frac{3}{2};\frac{x^2}{4}\right)}{\Gamma (i y+2)}-\sinh \left(x+\frac{\pi  y}{2}\right)\right)
\end{multline}
with the generalized hypergeometric function\footnote{See also: \url{http://functions.wolfram.com/HypergeometricFunctions/Hypergeometric1F2/02/} } $\,_1F_2(a_1;b_1,b_2;z)=\sum_{k=0}^\infty \frac{(a_1)_k z^k}{(b_1)_k (b_2)_k k!}$. 
The corresponding expression for $D_2$ follows directly using $D_2(\Omega_A,\Omega_B)=-C_2(\Omega_A,-\Omega_B)$.

Both the limit $\lim_{y\to0}C_2$ and $\lim_{x\to0}C_2$ of \eqref{eq:C2corr} exist, i.e., there exist closed expressions for zero-gap detectors. The former is an expression in terms of hypergeometric functions, while 
\begin{align}\label{eq:xlimit}
\lim_{x\to0} C_2=\frac\ii4 \text{sech}\left(\pi y/2\right).
\end{align}
This means that both in the limit of infinite acceleration $a\to\infty$, or equivalently when $\Omega_A,\Omega_B\to0$, the leading order signal strength approaches
$$
    |C_2|+|D_2|\to\frac{\lambda_A\lambda_B}2.
$$
It is interesting to note that  this limit can be approached in two mathematically equivalent ways which from a physical perspective appear rather different. 
On the one hand, the limit can be achieved by $\Omega\to0$, i.e., by diminishing the detector energy gap. 
On the other hand, the limit can equally be achieved by $a\to\infty$, i.e., by increasing the acceleration, which  also means that the minimal distance of $1/a$ between Alice and Bob at $t=0$ is decreased. 
In (3+1)-dimensional Minkowski spacetime these seemingly different physical effects have an identical impact on the leading order signal strength, because it only depends on the ratio of detector gaps and acceleration of the sender.
The leading order signal strength $|C_2|+|D_2|$ exhibits non-trivial features  for low $\Omega_B/a$
which appear on a smaller scale then resolved by Fig.~\ref{fig:3dplot}, but are highlighted in  Fig.~\ref{fig:detail}.


In (1+1)-dimensional Minkowski spacetime the integrals in $C_2$ and $D_2$ are easy to solve due to the simple form of the commutator \eqref{eq:comm11D}. The results are finite for arbitrary finite interaction times. However, the limit of infinite interaction times does not exist because of oscillatory contributions from signaling between timelike separated points (which we will discuss in the subsequent section). To suppress these oscillatory terms we introduce switching functions for Alice and Bob of the form
\begin{eqnarray}
\eta_d(\tau_d)=\ee{-|\tau_d|/\sigma_d}.
\end{eqnarray}
such that the   integrals converge for infinite interaction times. Then, after the integration is performed, we take the limit $\sigma_d\to \infty$ of infinite interaction times. Then \cite{jonsson_decoupling_2016}
\begin{eqnarray}
C_2=\lim_{\sigma_A,\sigma_B\to\infty} C_2(\sigma_A,\sigma_B)=-\lambda_A\lambda_B\frac{\left(-\frac{\ii\Omega_B}a\right)^{\ii\Omega_A/a}\Gamma\left(-\ii\Omega_A/a\right)}{2a\Omega_B},
\end{eqnarray}
which yields
\begin{eqnarray}
|C_2|+|D_2|=\frac{\lambda_A\lambda_B}{a\Omega_B} \cosh\left(\frac{\pi\Omega_A}{2a}\right) \sqrt{\frac{a\pi}{\Omega_A\sinh(\pi\Omega_A/a)} }.
\end{eqnarray}
The appearance of the overall factor $\lambda_A/a$ in front of this contribution suggests that in this scenario Alice's proper acceleration sets the scale in comparison to which the dimensionful coupling constant ought to be small in order for the perturbative analysis to be valid, i.e., $\lambda_A<<a$.

In (1+1)-dimensional Minkowski spacetime the limit of infinite acceleration exists. The maximum leading order signal strength across an acceleration horizon in (1+1)-dimensional Minkowski spacetime, achieved in this limit, is
\begin{eqnarray}
\lim_{a\to\infty}|C_2|+|D_2|=\frac{\lambda_A\lambda_B}{\Omega_A\Omega_B}.
\end{eqnarray}

\subsection{Timelike Signals}\label{sec:timelike}

The appearance of  signals propagating slower than the speed of light in massless fields may appear counter-intuitive, since they do not appear in (3+1)-dimensional Minkowski spacetime. However, this is an exception: Generically,  the Green function of a classical massless field, and thus the commutator of massless quantum fields, has support inside the future lightcone and timelike signaling is possible \cite{czapor_hadamards_2007,jonsson_information_2015,blasco_violation_2015,blasco_timelike_2016,simidzija_information_2017,jonsson_decoupling_2016}, as we saw above, \eqref{eq:comm11D} and \eqref{eq:comm21D}, is the case in (1+1)-dimensional and (2+1)-dimensional Minkowski spacetime. The latter two cases are particularly interesting because massless Klein-Gordon fields in one spactial dimension are realized in waveguides of superconducting circuits, and the two-dimensional case might be realizable, e.g., in graphene.

Here we show that for timelike signals, in contrast to null signals, the leading order signal strength does not depend on the sender and receiver being tuned into resonance. Instead, both the receiver and the detector only need to independently optimize their switching times with respect to their own detector energy gap. 

The reason for this is that once the receiver is timelike separated from the sender the boundaries of the time integrals in $C_2$ and $D_2$ are not interdependent any more: The upper boundary of the $t_2$-integral in \eqref{eq:C2tightbounds} is  always  the upper bound of the support of Alice's switching function.


Therefore, in contrast to null signals, timelike signals behave like  static remainders from Alice's interaction with the field. This remainder is imprinted into the field amplitude, and decays if the volume of the future lightcone increases.
To detect this static imprint, Bob does not need to tune his detector to Alice's parameters but can just locally pick the parameters which allow him to best measure the field's amplitude.

The lack of resonance effects between timelike separated detectors also means that the signal strength is bounded and cannot be increased by coupling the detectors to the field for longer times.

In (1+1)-dimensional Minkowski spacetime timelike separation between Alice and Bob even factorizes $C_2$ and $D_2$ into a product, because the commutator takes the constant value $\ii/2$ inside the future lightcone. Using $\chi_d(t)=\eta_d(\tau_d)\difffrac{\tau_d}{t}$, we obtain
\begin{eqnarray}
C_2 &=\lambda_A\lambda_B\integral{t_1}{}{} \integral{t_2}{}{t_1}\, \chi_A (t_2) \chi_B (t_1) \ee{\ii (\Omega_B  \tau_B(t_1) - \Omega_A \tau_A(t_2))} \frac\ii2 \nonumber\\
&= \frac\ii2\lambda_A\lambda_B \left( \integral{\tau_A} {}{}\, \eta_A ( \tau_A) \ee{ -\ii \Omega_A \tau_A} \right)  \left(\integral{\tau_B}{}{} \eta_B ( \tau_B) \ee{\ii \Omega_B  \tau_B}\right).
\end{eqnarray}
This shows that for strictly timelike separations of sender and receiver, $C_2$ is just the product of the individual switching functions' Fourier transforms, evaluated at the detectors' energy gaps. Most interestingly, the signal strength is independent of the detectors' motion and the separation between the detectors. It only depends on the switching of the interaction as a function of the detector proper times.

This form of $C_2$ implies that the leading order signal strength is maximized by sudden switching functions of the general form
\begin{eqnarray}
\eta_d(\tau)=\begin{cases} 1 \quad \text{if }\tau_0\leq\tau\leq\tau_0+\Delta\tau_d \\ 0 \quad \text{else}\end{cases}.
\end{eqnarray}
These sudden switching functions maximize the signal strength in the sense that modifying the switching by adding a  smooth ramp-up at the beginning of the interaction, and a  ramp-down at the end, generally decreases the absolute value of the Fourier transform.

With sudden  switching functions the leading order signal strength evaluates to
\begin{eqnarray}
|C_2|+|D_2|= \frac{4\lambda_A\lambda_B}{\Omega_A\Omega_B} \left| \sin\left(\Delta\tau_A\Omega_A/2\right) \sin\left( \Delta\tau_B \Omega_B/2\right)\right|.
\end{eqnarray}
So if both Alice and Bob couple  to the field for a proper time interval which corresponds to an integer and a half multiple of their own detector period, i.e., for $\Delta\tau_d=(n+1/2)2\pi/\Omega_d$, then the maximal leading order signal strength for timelike detectors in (1+1)-dimensional Minkowski spacetime is 
\begin{eqnarray}
|C_2|+|D_2|=\frac{4\lambda_A\lambda_B}{\Omega_A\Omega_B}.
\end{eqnarray}

\begin{figure}
\centering
\includegraphics[width=1\columnwidth,trim={0 6.5cm 0 7cm},clip]{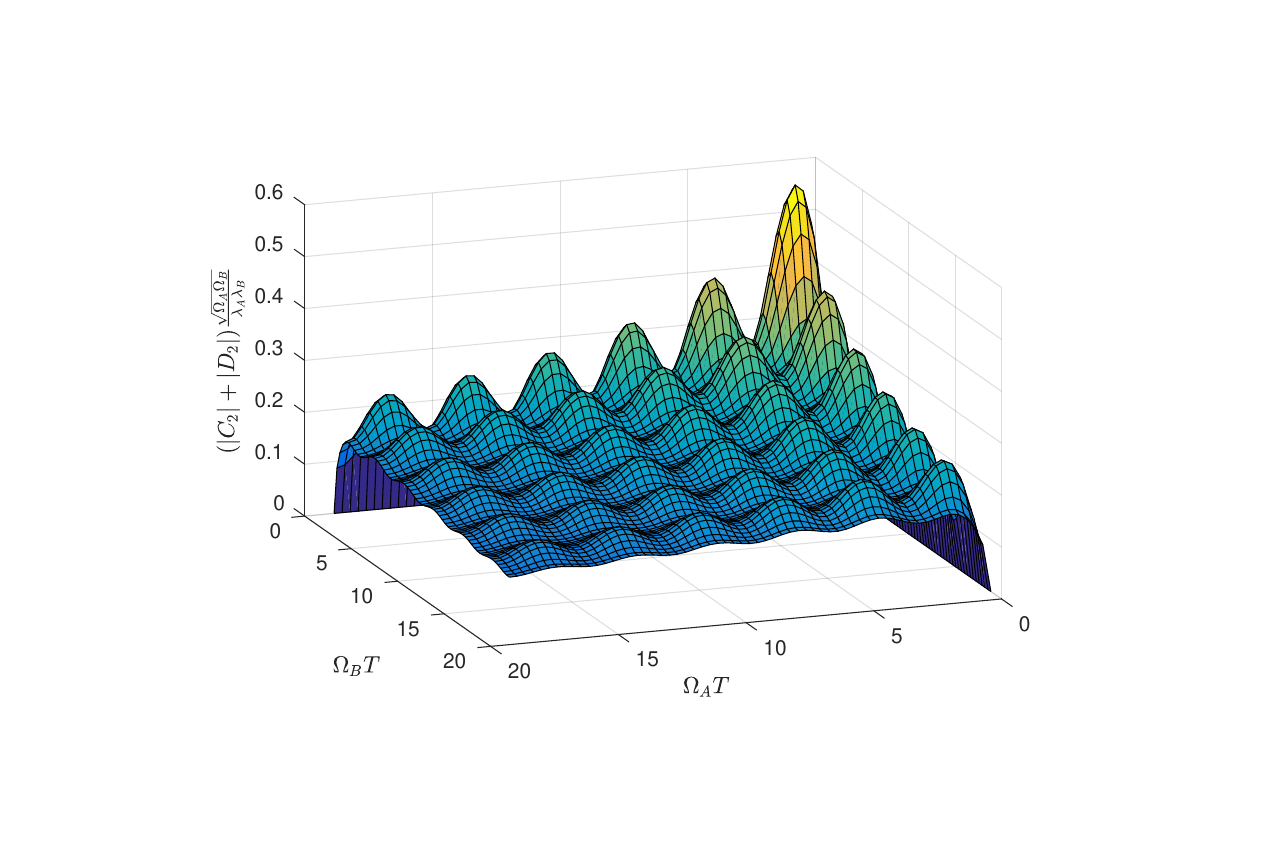}
\caption[signal strength for timelike signaling in 2+1D]{The leading order signal strength for timelike separated sender and receiver in 2+1D Minkowski spacetime for different  energy gaps $\Omega_A,\Omega_B$ of the detectors. The sender and receiver rest at the same position. The sender couples to the field for $t=0...2T$, and the receiver for $t=2.1 ... 4.1T$, with $T$ being a unit of time. 
The signal strength only requires the sender and receiver to individually choose optimal coupling parameters but, in contrast to null signals, does not benefit from resonance of the two detectors. 
Note that in 2+1D the coupling constant has mass dimension $[\lambda]=\frac12$, i.e., $\lambda/\sqrt{\Omega}$ plays the role of the perturbative parameter. Therefore, we plot the signaling strength in multiples of $\sqrt{\Omega_A \Omega_B}/(\lambda_A\lambda_B)$. (Figure adapted from \cite{jonsson_decoupling_2016}).
}
\label{fig:timelike2D}
\end{figure}

Also in (2+1)-dimensional Minkowski spacetime timelike signaling is possible because the commutator \eqref{eq:comm21D} has timelike support. However, here the strength of timelike signals decays with increasing timelike separation between sender and receiver because the commutator is proportional to $\propto 1/\sqrt{\Delta t^2-\Delta\bi{x}^2}$. Through numerical evaluations of the leading order signal strength we can confirm that it decays as
\begin{eqnarray}
|C_2|+|D_2|\sim\frac1{\Delta T}
\end{eqnarray}
for increasing timelike delays $\Delta T= T_1-T_A\to\infty$ between the switch-off of a resting sender and the switch-on of a resting, timelike separated receiver. This decay is slower than the decay of the energy content of the signal, which falls of with a higher power of $\Delta T$, as shown in \cite{jonsson_information_2016}. Therefore, the ratio of signal strength to transmitted energy of timelike signals grows unbounded, just as discussed for null signals above.

In particular, as shown in Figure \ref{fig:timelike2D}, also in (2+1)-dimensional Minkowski spacetime timelike separated senders and receivers need not tune their detectors into resonance but instead only need to individually pick the right amount of coupling time with respect to their detector energy gap.

\section{Conclusions}
The results discussed in this article complete the general analysis  of  signaling effects between two Unruh-DeWitt particle detectors when their interaction with the quantum field, through which they communicate, is treated within perturbation theory. 
We showed how the leading order signal strength can be maximized, and which classical channel capacity arises.

In particular, we found that the leading order signaling strength is given by the term $|C_2|+|D_2|$. This term has proven to be feasible to evaluate in many relativistic  scenarios, because it consists of a basic Fourier-type integral over the field commutator. 

The  leading order signal strength is not affected by the noise effects arising from the interaction of the receiver's detector and the vacuum fluctuations of the field. Because the field commutator is given by the classical Green function of the field.  Among the different measures of classical capacity considered, at leading order, only the Holevo capacity was affected by the field's vacuum fluctuations.

Here, we evaluated the leading order signal strength for detectors communicating via a massless field while moving relativistically through Minkowski spacetime of 1+1 to 3+1 dimensions. Here we observed the impact of the relativistic Doppler effect. Observers in inertial motion need to detune to account for the Doppler effect in order to maximize the signal strength. However, if either sender or receiver is uniformly accelerated the Doppler shift is infinitely large at early and at late times such that the signal strength across an acceleration horizon in Minkowski spacetime has an upper bound.

Both for null and for timelike signals, and in the different dimensional Minkowski spacetimes, we found that the signal strength decays slower than the energy density of the signal when the distance between sender and receiver is increased. This shows that the transmission of information through the amplitude of a massless field is not tied to the transmission of a minimum amount of energy along with the information from the sender to the receiver. Instead the propagation of information can decouple from the propagation of energy in the field.

The simple expression $|C_2|+|D_2|$, for the leading order signal strength, is general and applies also to massive fields and to curved spacetimes. A similar estimate for the signaling strength, which didn't yet include the optimal choice of initial states for Alice and Bob, was already used to investigate  timelike signals in expanding universes \cite{blasco_violation_2015,blasco_timelike_2016,simidzija_information_2017}. 

In future work, it will be interesting to study the signal strength in other curved spacetime scenarios, and to study information propagation in massive fields. Scenarios in Schwarzschild or Kerr black hole spacetimes are particularly interesting and can be treated using the results and methods  of \cite{casals_self-force_2013,yang_scalar_2014}.

Within the scope of this article we only discussed the transmission of classical information.
These results could be combined with previous results about the entanglement extraction from the field's vacuum state \cite{reznik_violating_2005,steeg_entangling_2009,cliche_vacuum_2011,salton_acceleration-assisted_2015,martin-martinez_entanglement_2014} to study quantum teleportation only by means of particle detectors, within the perturbative regime.

For other tasks, such as the direct coherent state transfer from one detector to another, non-perturbative interactions between the detectors and the field are necessary. So far non-perturbative solutions are only known for certain restricted types of couplings, such as  for zero-gap detectors \cite{landulfo_nonperturbative_2016}.
However, it is possible to achieve quantum capacity between particle detectors by combining these non-perturbative approaches with sequences of appropriate couplings between the detectors and the field \cite{kae}.

\ack
I am grateful to Achim Kempf and Eduardo Mart\'{i}n-Mart\'{i}nez for helpful discussions, and their support. I also thank Robert K\"onig for helpful discussions and for bringing \cite{berry_qubit_2005} to my attention. I acknowledge financial support from the Knut and Alice Wallenberg Foundation. I thank Marc Casals for bringing a mistake in Eq.~\eqref{eq:C2corr} in earlier versions of this article (and in \cite{jonsson_decoupling_2016}) to my attention.

\appendix
\section{Leading order single detector channel coefficients}\label{app:terms}
The leading order terms for the single detector coefficients of the channel in equation \eqref{eq:chanleadingorder} are \cite{jonsson_quantum_2014}
\begin{eqnarray}
\fl\quad&P_2= \lambda_B^2\integral{t_1}{}{}\integral{t_2}{}{} \chi_B(t_2) \chi_B(t_1) \ee{\ii \Omega_B(\tau_B(t_1)- \tau_B(t_2) )}  \exptval{\phi(x_B(t_2)) \phi(x_B(t_1))}\\
\fl\quad&Q_2=-\lambda_B^2 \integral{t_1}{}{}\integral{t_2}{}{} \chi_B(t_2) \chi_B(t_1) \ee{-\ii \Omega_B (\tau_B(t_1)- \tau_B(t_2) )}  \exptval{\phi(x_B(t_2)) \phi(x_B(t_1))}\\
\fl\quad&R_2=-\lambda_B^2 \integral{t_1}{}{}\integral{t_2}{}{t_1} \chi_B(t_2) \chi_B(t_1) \ee{\ii \Omega_B(\tau_B(t_1)- \tau_B(t_2) )} 2\Re \left[ \exptval{\phi(x_B(t_2)) \phi(x_B(t_1))}\right]\\
\fl\quad&S_2=\lambda_B^2 \integral{t_1}{}{}\integral{t_2}{}{} \chi_B(t_2) \chi_B(t_1) \ee{-\ii \Omega_B (\tau_B(t_1)+ \tau_B(t_2) )}  \exptval{\phi(x_B(t_2)) \phi(x_B(t_1))},
\end{eqnarray}
where, as before,  we absorbed the switching function and the time derivative of the detector proper time into $\chi_B(t)=\eta_B(\tau_B(t)) \difffrac{\tau_B(t)}{t}$.

\bibliographystyle{unsrt}
\bibliography{kae,Blochpicmovdec}
\end{document}